\documentclass[12pt]{article}
 \usepackage{amsmath,amssymb}

 \usepackage{epigraph}

\usepackage{etoolbox}

\makeatletter
\patchcmd{\epigraph}{\@epitext{#1}}{\itshape\@epitext{#1}}{}{}
\makeatother

 \newcommand{\be}{\begin{equation}}
 \newcommand{\ee}{\end{equation}}
 \newcommand{\bl}{\begin{equation}\begin{array}{ll}}
 \newcommand{\el}{\end{array}\end{equation}}
 \newcommand{\bll}{\begin{equation}\begin{array}{lll}}
 \newcommand{\bdm}{\begin{displaymath}}
 \newcommand{\edm}{\end{displaymath}}
 \def\bea{\begin{eqnarray}}
 \def\eea{\end{eqnarray}}
 \def\barr{\begin{array}}
 \def\earr{\end{array}}
 \newcommand{\bean}{\begin{eqnarray}}
 \newcommand{\eean}{\end{eqnarray}}
\def\p{\partial}
\def\dif{\partial}

\def\f{\varphi}
\def\ve{\varepsilon}
\def\ep{\epsilon}

 \def\La{\Lambda}
 \def\al{\alpha}
 \def\ga{\gamma}

\def\half{\frac{1}{2}}

\def\third{\frac{1}{3}}
\def\2third{\frac{2}{3}}
\def\4third{\frac{4}{3}}
\def\3quart{\frac{3}{4}}

\def\sixth{\frac{1}{6}}

\def\ra{\rightarrow}

\def\pr{\prime}

\def\ba{\bar{a}}

\def\bg{\bar{g}}
\def\bu{\bar{u}}
\def\bv{\bar{v}}
\def\bl{\bar{l}}

\def\bx{\bar{x}}

\def\bk{\bar{k}}
\def\bl{\bar{l}}

\def\bC{\bar{C}}
\def\bN{\bar{N}}
\def\bP{\bar{P}}

\def\bchi{\bar{\chi}}

\def\bxi{\bar{\xi}}
\def\bet{\bar{\eta}}

\def\cH{{\cal H}}
\def\cL{{\cal L}}

\def\ddx{\ddot{x}}
\def\dda{\ddot{a}}

\def\dvphi{\dot{\f}}

\def\ddpsi{\ddot{\psi}}

\def\dxi{\dot{\xi}}

\def\dsig{\dot{\sigma}}
\def\dpsi{\dot{\psi}}
\def\dal{\dot{\alpha}}
\def\ddal{\ddot{\alpha}}
\def\dbe{\dot{\beta}}
\def\ddbe{\ddot{\beta}}

\def\deta{\dot{\eta}}

\def\hep{\hat{\epsilon}}
\def\het{\hat{\eta}}

\def\hr{\hat{r}}
\def\hc{\hat{c}}
\def\hs{\hat{s}}

\def\pral{\alpha^{\prime}}

\def\prga{\gamma^{\prime}}
\def\prpsi{\psi^{\prime}}

\def\prl{l^{\prime}}

\def\prv{v^{\prime}}

\def\prx{x^{\prime}}
\def\pry{y^{\prime}}

\def\teta{\tilde{\eta}}

\def\tv{\tilde{v}}

\def\tg{\tilde{g}}

\def\tl{\tilde{l}}
\def\tpsi{\tilde{\psi}}

\usepackage{amsmath}

\begin{document}
\raggedbottom

 \title{{\bf On solving dynamical equations in general homogeneous isotropic cosmologies \\ with scalaron }}

 \author{A.T.~Filippov \thanks{Alexandre.Filippov@jinr.ru} \\
{\small \it {$^+$ Joint Institute for Nuclear Research, Dubna, Moscow
 Region RU-141980} }}

 \maketitle

 \begin{abstract}
  We study general gauge-dependent dynamical equations describing homogeneous isotropic cosmologies coupled to a scalar field $\psi$ (scalaron). For flat cosmologies ($k=0$), we analyze in detail the previously proposed gauge-independent equation describing the differential, $\chi(\alpha)\equiv\psi^\prime(\alpha)$, of the \emph{map of the metric $\alpha$ to the scalaron field $\psi$}, which is the main mathematical characteristic locally defining a `portrait' of a cosmology in the so-called `$\alpha$-version'. In a more habitual `$\psi$-version', the similar equation for the differential of the inverse  map, $\bar{\chi}(\psi)\equiv \chi^{-1}(\alpha)$, can be  solved asymptotically or for some `integrable' scalaron potentials $v(\psi)$. In the flat case, $\bar{\chi}(\psi)$ and $\chi(\alpha)$ satisfy the first-order differential equations depending only on the logarithmic derivative of the potential, $l(\psi) \equiv v^\prime(\psi)/v(\psi)$. Once we know a  general analytic solution for one of these $\chi$-functions, we can explicitly derive all characteristics of the cosmological model.

  In the $\alpha$-version, the \emph{whole dynamical system is integrable} for $k\neq 0$ and with any `$\alpha$-potential', $\bar{v}(\alpha)\equiv v[\psi(\alpha)]$, replacing $v(\psi)$. There is no \textit{a priori} relation between the two potentials before deriving $\chi(\alpha)$ or $\bar{\chi}(\psi)$, which implicitly depend on the potential itself, but relations between the two pictures can be found by asymptotic expansions or by \emph{inflationary perturbation theory}. We also consider alternative proposals -- to specify a particular cosmology by guessing one of its portraits and then finding (reconstructing) the corresponding potential from the solutions of the dynamical equations.

  The main subject of this paper is the mathematical structure of isotropic cosmologies, but some explicit applications of the results to a more rigorous treatment of the \emph{chaotic inflation} models and to their comparison with the \emph{ekpyrotic-bouncing} ones are  outlined in the frame of our `$\alpha$-formulation' of isotropic scalaron cosmologies. In particular, we establish an \emph{inflationary perturbation expansion} for $\chi(\alpha)$. When all the conditions for inflation are satisfied, which are: $v>0$, $k=0$, $\chi^2(\alpha) < 6$, and $\chi(\alpha)$ obeys a certain boundary (initial) condition at $\alpha\rightarrow -\infty$, the expansion is invariant under scaling of $v$ and its first terms give the standard \emph{inflationary parameters}, with higher-order corrections. When $v<0$ and $6\bar{\chi}^2 < 1$ our general approach can be applied to studies of   more complex \emph{ekpyrotic} solutions alternative to inflationary ones.

 \end{abstract}

 \newpage
 \bigskip \bigskip
 \tableofcontents

 \section{Introduction}
 In this paper we mostly study general Friedman-type isotropic cosmologies with one effectively massive scalar field (\emph{scalaron}, or, \emph{inflaton}), see, e.g., \cite{Mukhanov} - \cite{Rubakov}. First of all, we have in mind cosmological models pretending to describe a `pre-Big-Bang' evolution of the Universe in the frame of general relativity supplemented with a scalar field the exact nature of which is not important for us. The most successful models of this sort were discovered in papers \cite{Alstar}-\cite{StarLinde} in which there were proposed various models of inflationary cosmology.\footnote{A good supplement to its general  presentations in books \cite{Mukhanov} - \cite{Rubakov} might be the summary of main ideas in \cite{Linde90} and of more recent development in \cite{Linde-lec}-\cite{Martin}. For our notation see also Appendix~6.1.} However, at the moment, some alternative models are not completely excluded by the observational data and also attract attention (see, e.g., papers \cite{Mukhanov2}, \cite{Encycl},  \cite{Star-fast}-\cite{Star-bounce}, and many references therein).

 Here \emph{our aim is to spell out the mathematical structure of the general isotropic cosmologies}\footnote{We do not suppose that the space curvature parameter $k$ is zero, do not choose a special frame  (gauge), but for the moment assume the minimal scalaron coupling to gravity  and  arbitrary potential $v(\psi)$. Non-minimal cases can be considered by applying the well-known Weyl transformation of the metric.} and to study, in particular, gauge independent solutions of the cosmological dynamical systems in the $\al, \psi, t$-\emph{versions}. The last one, which we call the \emph{standard version}, is usually considered as a most useful in cosmological consideration. However, the dynamical equations are rather complex even in the case of vanishing curvature parameter $k$, when they can be analytically solved only for simplest potentials. For this reason, there was invented (after many `trials and errors') an intuitive `slow-roll' approximation which allowed one to construct the so-called \emph{chaotic inflation}, describing a very short but important interval in the evolution of homogeneous isotropic cosmologies \cite{Linde83}. This approach to cosmological evolution is rather visual and agrees with all well established cosmological data but it leaves aside the initial conditions and global picture of classical solutions, not speaking on existing viable alternatives to inflation. In our former search for exact integrals and approximations in non-isotropic cosmological and static reductions of spherically symmetric gravity coupled to scalars (see \cite{ATF1}-\cite{ATF14}), we had found  a different formulation of the relevant dynamical systems which we here develop and apply to general isotropic cosmologies including inflation.

 The first nontrivial example was the integrable model of gravity coupled to a massless scalaron studied in \cite{ATF1}. The idea is to construct the potentials, for which there exist additional integrals. Then the $\al$-dependence of the dilaton function $\beta$, which is the part of the higher-dimensional metric, can be derived by one integration. For the simplest model one can draw the $(\al, \beta)$-portrait of the static and cosmological solutions.\footnote{In this model the effective potential includes the electromagnetic term. When the scalaron coupling is minimal, the result can be written in terms of elementary functions. A topologically correct sketch of the portrait  derived in \cite{ATF1} (see also \cite{ATF13}) can be found in unpublished report \cite{ATFdraft}.} Like a phase portrait of a simple dynamical system, it consists of curves filling a domain in the $(\al,\beta)$-plane and having a few singularities which look like nodal or saddle points. We derive the curves of this portrait analytically but in general it may be considered as an object of differential topology.

 This seemingly artificial "looking for the lost keys under the street lamp" turned out to be useful both in constructing approximate solutions \cite{FM} and in finding new integrals of motion \cite{ATF13}-\cite{ATF14}. The main problem in anisotropic cosmology and in general static models is that we have three dynamical equations, two for the metric functions $\al(t)$ and $\beta(t)$ and one for the scalar matter $\psi(t)$, while we are granted only one integral -- the energy constraint. A symmetry of the potential sometimes provides us with one more integral but the third canonically commuting integral is a rarity, at least, in physically interesting systems.

 The structure of the spherically symmetric reduction, which is the two-dimensional field theory, allows one to find some integrable classes of models if we make strong simplifying assumptions about their potentials. For some multi-exponential potentials and the simplest (`minimal') coupling of scalars to gravity, there exist integrable systems related to Liouville or Toda-Liouville two-dimensional theories (see \cite{ATF1} - \cite{VDATF2}).\footnote{Such integrable models were, in particular, obtained in supergravity theories (see, e.g., \cite{Lidsey}-\cite{KL15}). Pure exponential potentials are interesting theoretically but were not so popular in cosmological applications.} For the one-dimensional cosmological reductions with one scalaron there might exist more integrable models. The simple exponential potential and a special potential with two exponential terms are integrable. But more interesting polynomial potentials are considered as not integrable and are usually studied in approximations that can be treated asymptotically or numerically.
 \emph{What must be the scalaron potential in the models of very early Universe is not completely understood.} In the `standard' model of chaotic inflation, the potential $v(\psi)$ proportional to $\psi^2$ was most popular but it may be replaced by more complex potentials. Strictly speaking, the models of the very early Universe, including the most successful inflationary models, do not yet form a completed theory, and they deserve deeper studies together with their alternatives.

 As a first step we propose to develop a fresh view of the mathematical structure of a fairly general homogeneous isotropic cosmologies. The next step will be to analyze them together with the (non-isotropic) static and wave sectors. The third step must be to study their interrelations in the frame of the two-dimensional theory of spherical gravity coupled to scalarons. This program had been attempted to partly implement in \cite{ATF2}-\cite{ATF13}, in the context of the multi-Liouville and Toda-Liouville integrable models, where the three sectors of the solutions were derived and classified. The difficult problem that was not really touched in that papers is the derivation and study of the solutions in presence of small perturbations. Here we make only \emph{the first step -- formulating different approaches to constructing approximate and exact solutions of cosmological dynamics with arbitrary potentials $v(\psi)$.}

 The main instrument is the \emph{gauge-independent $\chi$-equation} that was first introduced in \cite{ATF} for the case $k=0$. One of the main results of this paper is the derivation of some exact and approximate solutions of the equation for $\bchi(\psi)\equiv \al^\pr(\psi)$ and a fairly complete investigation of their asymptotic properties for large and small $\psi$ within  a physically important class of the potentials. To gain access to a most general picture we first introduce and discuss the dynamical equations in the arbitrary linear gauge ($c$-gauge) defined by the simplest choice of the Lagrange multiplier $e^\ga$ in terms of the `physical' isotropic metric $e^{2\al}$, namely, $\ga=-c\al$. Different gauge  choices are first illustrated by integrable models and special solutions on which we also illustrate the concept of portraits in cosmology, generalizing the phase portraits for the gravitational and matter subsystems and those proposed in Refs.~\cite{ATF1},~\cite{ATF13}.

 \emph{The most informative portraits in cosmology are the differentiable maps $\psi(\al)$ and $\al(\psi)$} that can be derived by integrating $\chi(\al) \equiv d\psi/d\al$ and $\bchi(\psi) \equiv d\al/d\psi$, respectively. The closed first-order differential equations for these functions can be written only when $k=0$. But even then we can analytically solve the equation for $\bchi(\psi)$ either with a few integrable potentials $v(\psi)$ or in the form of power-series or asymptotic expansions. In contrast, we show that \emph{the equation for $\chi^2(\al)$ can be explicitly solved if we formally replace $v(\psi)$ by $\bv(\al) \equiv v[\psi(\al)]$} which would be really possible if we knew $\chi(\al) = \psi^\pr(\al)$. Of course, we don't know it before solving the $\chi$-equation. Therefore, we should act as follows: take an arbitrary `potential' $\bv(\al)$, find the exact expression for $\chi(\al)$, derive its integral $\psi(\al)$, take the inverse map $\al(\psi)$, and thus eventually get $v(\psi) \equiv \bv[\al(\psi)]$. This is \emph{not a transformation} of a given function $\bv(\al)$ into a unique function $v[\psi(\al)]$ or vice versa. More precisely, in this chain   $\al(\psi)$ depends on two arbitrary integration parameters -- one is an initial or boundary condition $C_0$ for $\chi(\al)$ (see Section~4.3), the second, $\psi_0$ emerges in integrating $\chi(\al) = \psi^\pr(\al)$. In general it looks like a sort of \emph{`anchoring'} potentials in one version to some particular solutions in another one, i.e. the potential $v(\psi)$ will depend on $C_0$ (presumably, we can neglect $\psi_0$).~\footnote{See the simple examples and exact solutions in Appendices and in the main text.}

 Luckily, we live in a Unique Universe (the best or the worst) and $C_0$ must be fixed. In our  general $\al$-version formulation of the inflationary cosmology in Section~4.3 the condition $C_0=0$ gives the unique solution $\chi$ using which we freely jump from the exact $\al$-expressions to approximate (perturbative) ones in the $\psi$-version, with arbitrary potentials $\bv(\al)$ or $v(\psi)$. In this way we derive the \emph{general inflationary solutions}. We also discuss a more radical proposal: the potential is not a fundamental input in cosmology and we may venture instead to using one of the portraits. From the theoretical point of view, one of the best candidates to this role are the gravitational phase portrait $\dal(\al)$ and scalaron $\al$-portrait $\dpsi(\al)$.

 \section{Dynamical equations}
 In this paper, we consider the reduction of the two-dimensional theory  to equations describing isotropic cosmology and ignore other one-dimensional reductions studied in our previous work  \cite{ATF1} - \cite{VDATF2}. The procedure and notation is briefly described in Appendix~6.1. Note that the effective potential and the equations can also be directly derived from the spherically symmetric Einstein equations using the Lagrangian
 \be
  {\cal L}^{(4)} = \sqrt{-g} \,\, [ R^{(4)}  - v(\psi)
  - (\nabla \psi)^2 ] \,.
 \label{L1}
 \ee

 \subsection{Gauges and gauge independence}
 Supposing that the matter field depends only on time $\,t$ and applying the reduction procedure we find the \emph{effective cosmological Lagrangian} (see Appendix~6.1)
 \be
  \cL^{(2)} =
  e^{3\alpha - \gamma} (\dpsi^2 - 6 \dal^2) -
  e^{3\alpha + \gamma} v(\psi) - 6k e^{\alpha + \gamma}\,,
 \label{L2}
 \ee
 where now $\al, \beta , \ga$ depend on $t$ and $\beta = \al$. We see that here $e^\ga$ is the Lagrange multiplier related the parametrization invariance of the Lagrangian. We fix the \emph{class of possible gauges} by the linear condition $\ga + c\,\al =0$, where $c$ is an arbitrary real constant.

 This class of gauges includes: the \emph{standard} (`\emph{time}') gauge (S) $c=0,\,\ga = 0$; the \emph{Hamiltonian} gauge (H) $c=-3,\,\ga = 3\al $; the \emph{light-cone} gauge (LC) $c=-1, \,\ga = \al$. Here we first write and try to solve the main equations in the general linear gauge,  $\ga = -c\, \al$:
 \be
  \cL_c = e^{(3 + c)\al} (\dpsi^2 - 6 \dal^2) -
  e^{(3 - c)\al} v(\psi) - 6k e^{(1 - c)\al} \,.
  \label{L3}
 \ee

 First we vary $\cL_c$  in the gauge parameter $\ga$ and write \emph{the Hamiltonian (energy) constraint}:
 \be
  \cH_c \equiv \,\eta^2 - 6 \,\xi^2 +
  e^{-2c \al}\, v(\psi) + 6k \,e^{-2(1 + c)\al} \, = \,0\,; \qquad
  \eta \equiv \dpsi \,, \quad  \xi \equiv \dal \,,
 \label{H}
 \ee
 where we introduced the momentum-like variables $\eta\,, \xi$ that are more convenient than the canonical momenta $p_\psi = \p_{\,\eta} \cL_c\,$ and $\,p_\al = \p_{\,\xi} \cL_c\,$. Note that the canonical Hamiltonian,
 \be
  \cH_c^{\textrm{can}} = \, e^{-(3 + c)\,\al}\{p_{\psi}^2 /4 \,+\, v(\psi)\,e^{6\al} -\,p^2_\al/24\,+\,6\,k e^{4\al}\}, \quad p_{\psi}=2\eta e^{(3 + c)}, \,\,p_\al = 12\,\xi e^{(3 + c)},
 \label{HC}
 \ee
 coincides with the constraint $\cH_c$ only in the gauge $c=-3$, and we see that $\cH_c^{\textrm{can}}$ can never be split into dynamically independent scalaron and gravity parts (unlike $\cH_c$).

 \emph{The first-order equations} (equivalent to the canonical ones) are derived by varying Lagrangian (\ref{L3}) in $\psi$, $\al$, then replacing $\dpsi$, $\dal$  by $\eta$, $\xi$, and, finally, using constraint (\ref{H}):
 \be
  \dpsi = \eta\,, \quad \dal = \xi\,; \qquad 2\,\deta\,+\,2(3+c)\,\eta\,\xi \,+\, e^{-2c \al}\,\prv(\psi)\,=\, 0\,,
 \label{H1}
 \ee
 \be
  6\,\dxi\,+\,(3+c)\,\eta^2\,+\,c\,e^{-2c \al}\,v(\psi) \,+\,
 6k\,(1+c)\,e^{-2(1 + c)\al} \,=\, 0\,.
 \label{H2}
 \ee
 Expressing the $v$-term in the last equation using constraint (\ref{H}) we can rewrite it as
 \be
  2\,\dxi + \eta^2 +\, 2c\,\xi^2 +\, 2k e^{-2(1+c)\,\al} = 0\,,
 \label{H2c}
 \ee
 which is \emph{independent of $v(\psi)$ in any gauge}\footnote{In the S-gauge, $c=0$, it coincides with the previous equation (\ref{H2}).} and gives several interesting exact relations. For instance, taking in it $c=k=0$, we obtain below a very simple and useful equation (\ref{H2d}). If $c, k \geq 0$, we find the general exact inequality $\dxi \equiv \ddal \leq 0$. Similarly, from Eq.(\ref{H1}) in the gauge $c=-3$ we find $\deta \equiv \ddpsi \leq 0$ if $\prv(\psi)\geq 0$. One can derive from (\ref{H2c}) other interesting inequalities and exact relations for different values of $c,k$ and independent of $v(\psi)$. Moreover, one may \emph{consider equations} (\ref{H}), (\ref{H2c}) as the \emph{fundamental system} completely defining the cosmological solutions. The evident application is the reduction of the whole system to one differential equation for $\xi(\psi)$ in the case $k=0$, see (\ref{a4}).  Below we also propose a much less evident approach based on reinterpreting equations (\ref{H1})-(\ref{H2c}) as equations for $\xi^2(\al)$, $\eta^2(\al)$.

 These equations form our dynamical system of the first-order differential equations while $\cH_c(\eta,\psi;\,\xi, \al)$ is their first integral, which is constrained to zero as required by the parametrization invariance. The vanishing of $\cH_c$ is a separate condition if we wish to forget the origin of equations (\ref{H1})-(\ref{H2c}) from the Lagrangian theory (\ref{L2}). The standard formulation of cosmology uses the S-gauge and the second-order form of equation (\ref{H1}) supplemented by Eq.(\ref{H}). As we here use all equations (\ref{H})-(\ref{H2}) in different gauges and with different parameterizations of the dynamical variables it is important to clearly understand their interrelation.

 If equations (\ref{H1})-(\ref{H2}) are satisfied, $\cH_c$ is their integral of motion, i.e., $\dot{\cH_c}=0$ on their solutions. One can also find that, on solutions of equations (\ref{H1}) and (\ref{H2c}), the constraint  $\cH_c$ satisfies the equation $\dot{\cH_c} = -2c\,\xi\cH_c$. This means that their integral of motion is $e^{2c\al}\cH_c \equiv \bar{\cH}$, which coincides with the constraint if $c=0$ and thus $\cH_0$ is the integral in S-gauge. Now, suppose that we solve constraint (\ref{H}) and one of three equations (\ref{H1})-(\ref{H2c}). Then it is easy to check that the two other equations are satisfied by the solutions of the chosen pair.

 \subsubsection{Gauge-invariant equations and general remarks}
 Before we pass to more concrete problems we write the above equations in the gauge-independent form, which happens to formally coincide with the S-gauge. Indeed, equations (\ref{H})-(\ref{H2c}) are easily transformed into the $c\,$-independent form if we define the gauge invariant momentum-like variables $\bet$, and $\bxi\,$ and evolution parameter $\tau$,
 \be
  d\tau \equiv \,e^{-c\al}\,dt \,, \quad
  d/dt \equiv \,e^{-c\al}\,d/d\tau \,, \quad
  \eta\,\equiv e^{-c\al}\,\bet \,, \quad \xi\,\equiv e^{-c\al}\,\bxi \,,
 \label{H3}
 \ee
 in terms of which the main dynamical equations acquire the \emph{gauge-independent} form:
 \be
  \,d\psi/d\tau = \bet\,,\qquad 2\,d\bet/d\tau \,+\,6\,\bet\,\bxi \,+\, \prv(\psi) = 0\,,\,
 \label{H1a}
 \ee
 \be
  d\al/d\tau  = \bxi\,,\qquad 2\,d\bxi/d\tau \,+\,\bet^2 \,+\, 2 k\,e^{-2\al} = 0\,.
 \label{H2a}
 \ee
 \be
  \bar\cH\, \equiv \,e^{2c\al}\cH_c \,=\,\bet^2 - 6\,\bxi^2 + v(\psi) + 6k \,e^{-2\al} = 0\,.
 \label{Ha}
 \ee
 These equations are identical to Eqs.(\ref{H}), (\ref{H1}), (\ref{H2}) in the gauge $c=0$, if we omit bars and identify $d/d\tau$ with $d/dt$ (i.e., with the dot differentiation). In what follows we use this interpretation of (\ref{H1a})-(\ref{Ha}) without comments, if this will not lead to misunderstanding.

 Here we consider the simplest equations for spherically symmetric gravity coupled to a scalar field. As is well known, the simplification is defined by the isotropy condition (see (\ref{eq6}) in Appendix~6.1), which requires the equality $\beta(t) = \al(t)$ that is in fact an additional integral of the complete dynamical equations. To solve the reduced equations (\ref{H1})-(\ref{H2}) we need just one additional integral that commutes with the constraint. For instance, if $\prv(\psi) = 0$ we see that, in the H-gauge, $\eta$ is constant and the equations of motion can be explicitly solved. In anisotropic cosmologies,  there is an additional second-order equation for $\beta(t)$ and we need one more integral to solve the equations of motion, see the examples in \cite{ATF1}, \cite{ATF13}-\cite{ATF14}.

 The standard approach to cosmology usually deals with equations for time-dependent functions like (\ref{H1})-(\ref{H2}), or, with their second-order form. Also, the standard treatments uses the gauge $c=0$ ignoring all possible gauge transformations. Here, we mostly consider the general gauge non-invariant equations with $\psi$ or $\al$ treated as independent variables. As mentioned in Introduction, to solve the theory it is sufficient to derive either $\bchi(\psi) \equiv \pral(\psi)$ or $\chi(\al) \equiv \prpsi(\al)$. We show that, for $k=0$,  it is possible to find a first-order differential equation expressing $\bchi^\pr(\psi)$ in terms of a third-order polynomial of $\bchi$  with coefficients that are rational functions of $\psi$ if, for instance, $v(\psi)=e^{g\al}R(\psi)$, where $R(\psi)$ is a rational function.  The similar equation for $\chi(\al)$ can be completely solved in an unusual way that is briefly explained in Introduction and will be explicitly demonstrated below. To compare these two approaches we also discuss some exact and general asymptotic solutions of the $\bchi$ equation.

 \emph{The characteristic features of our approach} are the following. 1.~Unlike the standard practice, we use different gauges and \emph{versions} of dynamical equations, in which dynamical variables are parameterized not only by the gauge dependent parameter $t$, but also by invariant (`physical') parameters $\psi$ or $\al$. 2.~In addition to the standard phase portraits, $\eta(\psi)$ and $\xi(\al)$, we study `twisted' ones, $\eta(\al)$ and $\xi(\psi)$. 3.~The most important for us are the locally equivalent, invariant maps (portraits) $\psi(\al)$ and $\al(\psi)$ as well as their differentials (`sketches') $\chi(\al)= d\psi/d\al$ and $\bchi = d\al/d\psi$. 4.~We do not spent much efforts on search for classically integrable potentials; instead we try to find approaches to constructing exact and approximate solutions describing the portraits of wide enough classes of cosmologies.

 To illustrate these features we begin with a brief review of known integrable examples.

 \subsection{Simple examples from `upside-down' standpoint}
 With different forms of the equation and using different gauges one can easily solve several special cases. The most obvious integrable cases are: the constant (`cosmological') potential $v=v_0\equiv2\La$ and the more complex exponential potential, $v=v_0\exp{g\psi}$. They are important in the context of  our approach and will be considered from different viewpoints. In the first case, the last equation in (\ref{H1}) is easily solved in any gauge: noting that $\deta/\xi \equiv d\eta /d\al$ we find $\eta=\eta_0 \exp[-(3+c)\al]$. In the H-gauge, this integral of motion is simply $\eta = \eta_0$; in the LC-gauge we have the integral $e^{2\al}\eta = \eta_0$, which corresponds to the well-known integral of motion $\f\,\dpsi \equiv e^{2\beta}\eta = C_0$ of non-isotropic ($\beta \neq \al$) models of cosmologies and static states with the $\psi$-independent potentials.\footnote{The important property of this integrable dilaton gravity model is that for $C_0 \neq 0$ there is no horizon, which reappears when $C_0 =0$. If the potential depends on the dilaton field $\f$, it is in general not integrable. For integrable potentials $v(\f)$ we derived two-dimensional portraits that are the systems of curves in the plane ($h \equiv e^{2\beta}; \f$),  which essentially depend on one parameter $C_0\,$, and look like a phase portrait (\cite{ATF1}, \cite{ATF13}).} Now, having the explicit expression for $\eta(\al)$ we can derive $\xi^2 \equiv \dal^2$ from the constraint (\ref{H}) and thus find $t_c = \int{d\al/\dal(\al)}$ with $\dal(\al)$ defined by:
 \be
  6\dal^2 e^{2c\al} = v_0 + 6k\,e^{-2\al} + \eta_0^2\,e^{-6\al}\,.
 \label{Hc2}
 \ee
 This is the gravitational phase portrait $\dal(\al)$ of the cosmology with the constant cosmological potential. To derive the $\psi(\al)$ portrait is easiest in the H-gauge: when $c=-3$, we have $\psi = \eta_0 \,(t-t_0)$ and thus the gauge-independent portrait $\psi(\al)$ will be found if we derive $t(\al)$ in the same gauge.  The explicit general expression for $t(\al)$ can be obtained from (\ref{Hc2}) but for simplicity we write the final result for  $k=0$, $v_0 >0$ (for $v_0<0$, $\,\sinh^2 \mapsto \cosh^2$):
 \be
  \eta_0^2 \,e^{-6\al} =\,v_0\,\sinh^2[\sqrt{3/2}\,\eta_0\,(t-t_0)]\,=\,v_0\, \sinh^2[\sqrt{3/2}\,(\psi-\psi_0)]\,.
 \label{Hc3}
 \ee
 Like the portraits of simple  integrable dilaton gravity models discussed in \cite{ATF1}, \cite{ATF13}, this portrait essentially depends on one free dimensionless integration parameter $\eta_0^2 /v_0$ but has no interesting singularities except the usual ones at $\al\rightarrow \pm \infty$, $\psi \rightarrow \infty$, $\psi \rightarrow \psi_0$.

 Naturally, the model with the constant cosmological potential can be simply solved and analyzed in any gauge. The search for integrability of more complex cosmologies can be simplified by other gauge choices.
 Though the S-gauge is preferred by cosmologists, the H and LC gauges are often more convenient. The LC-gauge is indispensable in studies of unified description of static and cosmological models and it most directly relates them to higher dimensional theories (see, e.g., \cite{ATF1}-\cite{ATF2}, \cite{ATFp}-\cite{ATF14} and references therein). The H-gauge was extensively used in our searches for integrals and integrability of these models.

 \subsubsection{Solutions of equations with exponential potentials}
 Especially interesting and simple example of application of the H-gauge, $c=-3$, is the $k=0$ case. Then the Lagrangian and second-order equations of motion are
 \be
  \cL_c = \dpsi^2 - 6\dal^2 - e^{6\al} v(\psi) \,; \qquad
  2\,\ddpsi\,+\, e^{6\al}\,\prv(\psi)\,=\, 0\,,\quad
  2\,\ddal\,-\,e^{6\al}\,v(\psi) \,=\, 0\,.
 \label{He1}
 \ee
  If $v=v_0 e^{g \psi}$, there exists the obvious integral of motion $\dpsi\,+\,g\,\dal = C_0$ and, as is well known, the system is integrable. To explicitly integrate it we define $\f \equiv g \psi + 6\al$, which satisfies the Liouville equation  having the integral $\dvphi^2 +2\tv_0 e^\f = C^2_1$, where $\tv_0 \equiv v_0(g^2-6)/2$. It follows that the complete solution for the exponential potential can be written as:
 \be
  \psi + g \al = C_0 (t-t_0)\,, \quad e^{-(g\psi\,+\,6\,\al)} = \tv_0 C_1^{-2}\{1 + \ve\cosh[C_1(t-t_1)]\}\,; \qquad \ve \equiv \tv_0/|\tv_0|\,.
 \label{He1a}
 \ee
 Eliminating $t$ we find the relation between $\al$ and $\psi$ that implicitly define the portrait $\psi(\al)$:
 \be
  e^{-(g\psi\,+\,6\,\al)} = 2\tv_0 C_1^{-2}\cosh^2\{C_1[\psi + g \al + C_0(t_0-t_1)]/2C_0\}\,, \quad \ve=+1\,.
 \label{He1b}
 \ee
 It is easier to find from (\ref{He1a}) the explicit expressions for $\psi(t)$ and $\al(t)$. Deriving the inverse of $\al(t)$ we can find the `explicit' representation for $\psi(\al)$, but it seems that in this simple case the `parameterized portrait' ($\al(t)$, $\psi(t))$ is more convenient for cosmological applications.\footnote{The explicit representations for $\psi(\al)$ and $\al(\psi)$ can easily be derived in the asymptotic region $t\rightarrow \infty$.}

 A more interesting and realistic model described by Eqs.(\ref{He1}) is proposed in \cite{ATF14}. We remind that the potential $v=v_1\,e^{g_1\psi} +\, v_2\,e^{g_2\psi}$ is integrable when $v_i$ are arbitrary and $g_1 g_2 = 6$. Indeed, defining new fields $\psi_1, \psi_2$  by the pseudo-orthogonal transformation
 \be
  \sqrt{6}\, \al \equiv \hc \psi_1\,-\, \hs\psi_2\,, \quad \psi \equiv -\hs \psi_1\,+\,\hc\psi_2\,; \qquad \hc \equiv \cosh\theta \,,\quad \hs\equiv \sinh\theta \,,
 \label{He2}
 \ee
 we find that the kinetic Lagrangian is $\,-\dpsi_1^2 + \dpsi_2^2\,$ and, if $ g_1 =\,\sqrt{6}\tanh \theta , \,g_2 = 6/ g_1$, then
 \be
  6\al + g_i \psi = \tg_i \,\psi_i\,,\qquad \tg_1 = \sqrt{6}/\cosh\theta\,, \quad \tg_2 = \sqrt{6}/\sinh\theta \,.
 \label{He3}
 \ee
 The Lagrangian in (\ref{He1}) then describes two explicitly solvable Liouville models:
 \be
  \cL_c = -\dpsi_1^2 +\, \dpsi_2^2 + v_1 \,e^{\,\tg_1 \psi_1} +
  \,v_2 \,e^{\,\tg_2 \psi_2}\,.
 \label{He4}
 \ee
 As the parameters $v_1, v_2$ are arbitrary, $\tg_1 \tg_2>0$, and $\theta$ may be negative or positive, it is possible to construct cosmologically interesting models by choosing for them proper values. In particular, the potential $v(\psi)$ may qualitatively resemble inflationary potentials discussed in some models if we take $v_1 = -v_2 > 0$ and cleverly choose $\theta$ in the interval $(-\infty, +\infty)$.

 The formal solution of this model is simple: we know $\psi_i(t)$ and thus easily find $\al(t)$ and $\psi(t)$. Then, eliminating $t$ we  can derive the $\al(\psi)$ portrait of the system. This looks very simple but details are in fact cumbersome because, in general, $\ve_i$ and $C_i$ defining the solutions $\psi_i(t)$ (by two relations like the second equation in (\ref{He1a})) may be positive or negative and the inverse functions $t(\psi_i)$ are involved. All these `subtleties' are essential in cosmological applications. For these reasons we think that this `bi-Liouville' model undoubtedly deserves a detailed investigation that lies completely outside the scope of this paper.\footnote{More complex exponential integrable models with one and more scalar fields can be found in \cite{ATF2}-\cite{VDATF2}.}

 \subsubsection{Note on independence from potentials}
 The above simple examples demonstrate that it may be useful to replace the time variable by the more `physical' variable $\al$. We also can (and will) use the scalaron variable $\psi$, but $\al$ plays a special role because there exists equation (\ref{H2c}) independent of the potential $v(\psi)$.\footnote{A central idea of this paper is to look for the portrait of cosmology, $\psi(\al)$ or $\al(\psi)$, in terms of complementary variables $\psi$, $\al$. Thus it is quite natural to consider the dynamical variables as depending on one of these variables instead of the evolution (`time') parameter. In addition, we also discuss the standard phase portraits, $\xi(\al) \equiv \dal(\al)$, $\eta(\psi) \equiv \dpsi(\psi)$, as well as the `twisted' portraits, $\eta(\al) \equiv \dpsi(\al)$, $\xi(\psi) \equiv \dal(\psi)$.}

 This equation can be rewritten in a very compact form:
 \be
  \frac{dz}{d\al} + \bet^2 = 0\,; \qquad z \equiv \bxi^2 -k e^{-2\al}\,, \qquad \bxi \equiv \xi e^{c\al}\,, \quad \bet \equiv \eta e^{c\al}\,,
 \label{H2d}
 \ee
 where $-6z$ is the gravitational part of the gauge invariant  Hamiltonian constraint, and $(\bet^2 + v)$ is its scalaron part. The solution of equation (\ref{H2d}) expresses $\bxi^2$ in terms of $\bet^2$,\footnote{Here and below we often omit $d\al, d\psi, dx$, when possible.}
 \be
  \bxi^2(\al) =  ke^{-2\al} +\, z(\al_+)  + \int_{\al}^{\al_{+}} \bet^2(\al)\,,
 \label{H2e}
 \ee
 where $z(\al_+)$ is an arbitrary constant, $-\infty < z(\al_+) < \infty$.
 This is a nontrivial representation for the gravitational kinetic energy, $\dal^2$, in terms of the scalaron kinetic energy, $\dpsi^2$. It may be useful for analyzing the gravitational phase portrait $\dal(\al)$. More generally, one may find  entertaining the idea to use $\eta^2(\al)$ instead of the potential $v'(\psi)$. This resembles the simplest canonical  transformation of $p$ into $-q$ and will be discussed in some detail below.

 Possibly, a \emph{better idea is to take as an input} $\bxi(\al) \equiv \dal(\al)$ and derive $\bet(\al) \equiv \dpsi(\al)$ from Eq.(\ref{H2d}). In cosmological considerations, $\dal(\tau)$ is called the \emph{Hubble parameter} and its dependence on $\al$ is more or less understood in simple cosmological models and can be to some extent compared with observational data. It is possible to make reasonable guesses on it in various models of inflation, of bouncing, etc. From the dynamical point of view it is the phase portrait of the gravitational subsystem which in the domain `before Big Bang' is the main object of interest. We have no \emph{direct} observational information on this object but analyzing extensive literature on inflation and other scenarios of the very early cosmology one can develop intuition of what can be most plausible guesses for $\dal(\al)$. In any case, if we knew by any means one of the functions $\xi(\al)$ or $\eta(\al)$, we could find the second function from equation (\ref{H2d}) and then derive from constraint (\ref{H}) the potential itself as a function of $\al$. To `transform' it to $v(\psi)$ we must find $\chi(\al)$, as discussed below.

 Although the scalaron potential is a useful and convenient thing, we know about it much less than, say, on the Hubble parameter (not even speaking on the existence of the scalaron itself). Thus attempts to avoid using  $v(\psi)$ as an input in cosmological models look quite natural and this can also be done in the frame of the standard version. Then one of natural theoretical inputs may be the scalaron (inflaton) phase  portrait $\dpsi(\psi) \equiv \eta(\psi)$ that in turn can be found from some simplifying ansatz defining a concrete model (see, for example, the derivation of the potential in the model of `fast-roll inflation' in Ref.\cite{Star-fast}).

 \subsection{Equations for $\chi(\al)$, $\bchi(\psi)$ and their main properties}
 Now we suggest to forget, for some time, the brilliant original treatments of the isotropic cosmology and try to have a fresh look at its mathematical structure. My first attempt of this sort was made in \cite{ATF} (see also \cite{ATFn}, \cite{ATFs}) with the aim to understand cosmology of the affine generalization of Einstein`s gravity. In that  paper I have actually proposed (in a somewhat misleading notation gauge independent equation (33) for $\chi \equiv d\psi / d\al$, regarded as a function of $\psi$. We return to this equation later but first elaborate notation and try to explain more precisely our plan. The principal goal of this paper is to find the analytic portrait of cosmology in terms of either $\psi(\al)$  or $\al(\psi)$ and thus our main objects are\footnote{In this Subsection we use the standard gauge $c=0$ but omit bars over $\xi,\eta$.}:
 \be
  \chi(\al) \equiv d\psi / d\al \equiv \dpsi / \dal \equiv \eta / \xi \,, \quad\,\, \bchi(\psi) \equiv d\al / d\psi \equiv \dal / \dpsi \equiv \xi / \eta \,; \qquad \chi(\al)\,\bchi(\psi) = 1\,.
 \label{d1}
 \ee
 Evidently, the last relation is identity if we put into it $\psi(\al)$  or $\al(\psi)$.

 To derive the equations for $\chi(\al)\,$ and $\bchi(\psi)$ from (\ref{H1})-(\ref{H2}) we must use the transformation from the independent variable $\tau$ to $\al$ or $\psi$ as well as transformations between $\al$ and $\psi$ that we can easily perform with the help of the evident identities (implicitly applied in Section~2.2):
 \be
  \frac{d}{d\tau} =\, \dal \frac{d}{d\al} =\, \xi \frac{d}{d\al} =\, \dpsi \frac{d}{d\psi} =\, \eta \frac{d}{d\psi} \,\,;    \quad  \frac{d}{d\psi} =\, \bchi(\psi) \frac{d}{d\al}\,;    \quad \frac{d}{d\al} =\, \chi(\al)   \frac{d}{d\psi} \,\,.
 \label{d2}
 \ee
 Using the above equations we first derive the gauge independent relation,  \be
 \frac{2}{\xi}\frac{d\chi}{d\tau}  = \frac{\xi \deta -\dxi \eta}{\xi^3}  = (\chi^2 - 6) [\,\chi +\,l^{\pr}(\psi)\,] + \frac{2k}{\xi^2}\, e^{-2\,\al} [\,\chi +3\,l^{\pr}(\psi) \,]\,,\quad l^{\pr}(\psi) \equiv \frac{v^\pr(\psi)} {v(\psi)}\,,
 \label{d3}
 \ee
 which  generalizes Eq.(33) of \cite{ATF} but is not a closed differential equation for $\chi$, even if we replace $\xi\,d\tau$ by $d\al\,$ or by $\bchi\,d\psi$. In the first case, we find that this expression actually coincides with Eq.(33) of \cite{ATF}, when $k = 0$. It can be formally written as the well-defined equation for $\chi^2(\al)$ if we multiply it by $\chi(\al)$ and then apply the last relation of (\ref{d2}) allowing to formally consider $l^\pr(\psi)$ as a function of $\al\,$. Thus defining $\chi(\al) dl/d\psi \equiv d\bl/d\al = \bl^\pr(\al)$, we have
 \be
  \frac{d\chi^2}{d\al} = \bigl(\chi^2 - 6\bigr) \bigl(\chi^2 + \bl^{\pr}(\al)  \bigr) \,, \qquad \chi\equiv \psi^\pr(\al) \,,
  \label{d5}
  \ee
 which can be elementary solved for any $\bl^\pr(\al)$. In the second case, we have the closed Abel equation  for $\bchi(\psi)$, which is solvable just for a few potentials,
 \be
  2\,\frac{d\bchi}{d\psi} = \bigl(6\,\bchi^2 - 1\bigr) \bigl(1 +\, \bchi \, l^{\prime}(\psi)  \bigr)\,, \qquad \bchi \equiv \al^\pr(\psi)\,.
 \label{d9}
 \ee
 Equation (\ref{d5}) can be rewritten as the nonlinear second-order equation for $\psi(\al)$ if we recall that $\bl^{\pr}(\al)\equiv \chi(\al)\,l^\pr(\psi)$ and divide it by $\chi\,$. Similarly, (\ref{d9}) may be treated as the second-order equation for $\al(\psi)$ by recalling $l^\pr(\psi) \equiv \bchi(\psi)\,\bl^\pr(\al)$. In either transformations we should suppose that $l(\psi)$ or $\bl(\al)$ is known and thus the first case looks preferable. We do not study these second-order equations and in what follows concentrate on (\ref{d5}) and (\ref{d9}). As $v(\psi)$ is a more familiar thing, we begin our study with investigating solutions of (\ref{d9}).

 Note that constraint (\ref{H}) gives, among other things, the \emph{important inequalities} for $\chi, \bchi$:
 \be
  \chi^2 < 6\,, \,\, 6\bchi^2 > 1\,,\,\,\,\, \textrm{if} \,\,\, v> 0\,, \,\,k\geq 0\,; \qquad \chi^2 > 6\,, \,\, 6\,\bchi^2 < 1\,, \,\,\,\, \textrm{if} \,\,\, v < 0\,,\,\,k\leq 0\,.
 \label{d5b}
 \ee
 These restrictions on $v$ and $k$ are often used in cosmological considerations. If they are not realized, the behavior of $\chi$-functions becomes much more intricate, and we usually adhere to them. When $\chi^2 \ll 6,\,\, 6\bchi^2 \gg 1$, there may exist inflationary solutions, which we discuss in Section~4.3. For negative potentials, there exist viable alternative cosmological scenarios including a hot (ekpyrotic) compression followed by a bounce.

 In $k=0$ case the equation for $\chi(\al)$ (or, $\bchi(\psi)$) define the portrait $\psi(\al)$ (or, $\al(\psi)$) that can derived by directly integrating the solution. Moreover, once we know one of these functions we can in principle derive the complete solution of the cosmological equations. This is obvious when $k=0$ but can be generalized to $k\neq 0$, if we somehow find the general $\chi$. Then, taking in Eq.(\ref{H2d}) $\bet^2 \equiv \chi^2  \bxi^2$ we obtain the linear equation for $\xi^2$ and so get:
 \be
  \bxi^2(\al) \,=\, e^{-\int\chi^2 (\al)} \bigl[C_0 - 2k\int e^{-2\al + \int\chi^2 (\al)\,} \bigr] \,,
 \label{d4a}
 \ee
 expressing the Hubble function in terms of $\chi^2(\al)$. By changing the integration variable and remembering that $\chi^2(\al)\,d\al = \bchi^{-1}(\psi)\,d\psi$,  $\al^\pr(\psi) \equiv \bchi(\psi)$  we find $\bxi^2$ in the $\psi$-version:
 \be
  \bxi^2(\psi) \,=\, e^{-\int\bchi^{-1} (\psi)}\bigl[C_0 - 2k\int e^{\int [\bchi^{-1}(\psi)-2\bchi (\psi)\,]} \bigr]\,.
 \label{d4b}
 \ee
 In  applications of these formulas one has to carefully define the limits of integrations and the arbitrary constants to guarantee the positivity of $\xi^2$. As a first application we note that by using (\ref{d4a}), (\ref{d4b}), (\ref{Ha}) one can derive the potentials $\bv(\al)$ and $v(\psi)$ in terms of $\chi$ or $\bchi$:
 \be
  \bv(\al) = [\,6-\chi^2(\al)\,]\,\xi^2(\al)- 6k \,e^{-2\al} =\,  [\,6-\bchi^{-2}(\psi)\,]\,\xi^2(\psi) - 6k \,e^{-2\al(\psi)} = v(\psi)\,.
 \label{d4ab}
 \ee
 Also note that Eq.~(\ref{d4b}) can be used in iterations involving $\bchi(\psi)$ and $\xi$ or $\eta$, see Section~3.1.

 Equations (\ref{d4a})-(\ref{d4ab}) demonstrate (for arbitrary $k$) the \emph{most important property} of the functions $\chi, \bchi$ -- once we know one of them, we can derive not only the portrait $(\al, \psi)$ but all characteristics of the cosmological solutions: $\tau(\al)$, $\tau(\psi)$, $v(\psi)$, $v[\psi(\al)]$, gravitational and scalaron energies. Even more interesting for applications may be  the expression for $\chi^2(\al)$ in terms of the \emph{Hubble parameter as a function of the metric}, $H(\al) \equiv \bxi(\al)$.
 It can be obtained by rewriting equation (\ref{H2d}) that was also used for deriving Eq.(\ref{d4a}) (recall that $d\al \equiv Hd\tau$):
 \be
  \chi^2=-2\biggl(\frac{d\ln{H}}{d\al}+\frac{k\,e^{-2\al}}{H^2}\biggr)=
  -\frac{2}{H^2}\biggl(\frac{dH}{d\tau}+k\,e^{-2\al}\biggr)\,,
 \label{d4c}
 \ee
  This simple `inverse' formula \emph{allows us to reconstruct the complete solution of all dynamical equations and the potential}
 in the interval of $\al$ or $\tau$ in which we know $H$.

 It is worth noting that equations (\ref{d3})-(\ref{d9}) have an unusual property of being dependent on the logarithmic derivative, $l^{\prime}(\psi) \equiv [\ln v(\psi)]^{\prime}$, not on the potential itself. This property is shared by  the differential equations both for $\chi(\al)$ and $\bchi(\psi)$. This means that apparently small deformations of the potential $v(\psi)$ may result in significant deformations of the $\chi$-maps. For instance, if $v = \ve + \psi^n$, $n>1$, then for $\psi \rightarrow 0$ the `potential' $l^{\prime}(\psi)$ vanishes if $\ve \neq 0$, but it is singular for $\ve = 0$. On the other hand, the potentials $\psi^n$ give very similar $\chi$-maps as $\prl = n/\psi$. In the standard approach, $n=2$ looks  preferable as it gives, in the S-gauge, a linear equation for $\psi$, which however depends also on $\dal \equiv \xi$ and is not a closed equation for $\psi$ so far as $\ddal\neq 0$. Note also that the Higgs-type inflationary potentials may produce finite-range singularities $\sim 1/(\psi-\psi_0)$ that can be dealt with similarly to $n/\psi$ ones.

 A more fundamental problem is what to do with cosmologies having nontrivial non-integrable potentials and $k\neq 0$. It is evident that in this case the $\chi$-equations (\ref{d3}) are not simpler that the complete system (\ref{H})-(\ref{H2c}). However, basing on the structure of solutions of simple integrable cosmologies we propose here a different approach to $k=0$ equations for $\chi(\al)$ that allows us to find its simple analytical solution for apparently `general potential' $\bv(\al) \equiv v[\psi(\al)]$. Here $\psi(\al)$ is at first an unknown function that is to be derived \emph{after} solving the $\chi(\al)$-equation depending on $\bv(\al)$, i.e., $\psi(\al) = \int d\al\,\chi(\al)$.

 Indeed, suppose that we have derived $\chi(\al)$ corresponding to the potential $\bv(\al)$ and wish to find the potential $v(\psi)$ using the definition of $\bl^\pr(\al)$,
 \be
  \bl^\pr(\al) = \frac{d\ln\bv(\al)}{d\al} \,=\, \frac{d\psi}{d\al}\,\,\frac{d\ln v(\psi)}{d\psi} = \chi(\al)\, l^\pr(\psi)\,.
 \label{d5a}
 \ee
 Once we know $\chi(\al)$ and, correspondingly $\psi(\al)$, we can in principle derive the inverse function $\al(\psi)$. Replacing in (\ref{d5a}) $\al$ by $\al(\psi)$ we find $l^\pr(\psi) = \bl^\pr[\al(\psi)]/ \chi[\al(\psi)]$ and determine $v(\psi)$ up to a constant multiplier. At first sight, by writing $\bv(\al) \equiv v[\psi(\al)]$ we fix this problem but then there remains one constant that enumerates the curves $\psi(\al)$ corresponding to physically different solutions. It looks as if the solutions corresponding to the same $\psi$-potential may correspond to a one-parameter family of $\al$-potentials and vice versa. One can hope to better understand this  relation by further studies of the integrable examples given in Section~2.2 and Appendix~6.3 as well as of special solutions discussed in Appendix~6.2.

 \subsubsection{On what is the solution}
 Before turning to studies of approximate and exact solutions of the dynamical equations let us formulate what we call \emph{the solution} of our problem. In the Liouville sense, this means that the complete solution ($\al(t)$, $\psi(t)$) can be formally expressed in terms of integrals and derivatives of the potential, up to some functional inversions (at best, one can find explicit expressions for ($t(\al)$, $t(\psi)$), for $\al(\psi)$, or for $\psi(\al)$). In most favorable cases, we can find a complete phase portrait of the solution, say, the dependence of $\dal$ on $\al$ (or, $\dpsi$ on $\psi$) with calculable asymptotic behavior near singularities. This is possible in few cases. More often one can derive a differential equation for one function, like $\bchi(\psi)$, which expresses its first derivative, $\bchi^\pr(\psi)$, as a rational function of $\bchi$ and $\psi$.

 A cosmologically relevant example is the much studied Emden-Fowler type equations (see, e.g., \cite{Ed}-\cite{Bel}, \cite{ATF14}). To illustrate their relation to cosmological models let us consider equation (\ref{H1a}) for $\psi(\tau)$ derived in the $c=0$ gauge, which is usually written as:
 \be
  \ddpsi\,+\,3\,\dal\dpsi\,+\,v^\pr(\psi)/2 = 0\,,\qquad \dpsi = \eta\,,\quad \dal = \xi\,.
 \label{EF}
 \ee
 If we suppose that $\dal$ is a constant, $\dal=\xi_0\,$, and $v = a\psi^2 - b\psi^p$, where $p$ is a rational number, we obtain the Emden-Fowler equation. Actually, in cosmology described by equations (\ref{H1a})-(\ref{Ha}), the only possibility to have a nontrivial solution of equation (\ref{EF}) with the constant Hubble parameter $\dal\equiv\xi_0$ is taking $k\leq 0$ and $v = v_0 + a\psi^2$, so that it becomes linear.\footnote{See equations (\ref{c1})-(\ref{c3}) in Appendix~6.3,  where we also discuss further examples of the potentials for which the system (\ref{H1a})-(\ref{Ha}) with $\dxi=C_0\,$ or $\deta=C_0\,$ can be solved.} The general Emden-Fowler equation is usually rewritten in the form of a first-order nonlinear equation, e.g., $\eta^\pr(\psi) + 3\,\xi_0 + v^\pr(\psi)/2\eta=0$, which is the phase portrait of equation (\ref{EF}). Even with simple rational potentials, it cannot be solved analytically. Nevertheless, we better avoid calling it non-integrable. Indeed, the behavior of their solutions for rational $p$ was analyzed in great detail, including their exact asymptotic. More recently, a few significant results were also obtained on their classical integrability (see \cite{Ber}, \cite{EFint}).

 This example demonstrates why a general equation like $z^\pr(x) = R(x,z)$ with simple rational function $R(x,z)$ may be regarded as essentially  integrable: 1.~The topological portrait of the solutions can be found as far as we can derive the zeroes and poles of $R$ outside of which the solutions are locally analytic;  2.~According to Hardy's theorem \cite{Hardy}-\cite{Fowler2}, the possible asymptotic behavior of real continuous solutions is either $z\sim a x^b e^{P(x)}$ or $z\sim a x^b (\ln z)^{1/n}$, where $n$ is an integer, $P(x)$ - a polynomial. 3.~It follows that in simple cases we can derive approximate analytic behavior of the solutions in the (x,z)-plane, including singular points.

 Unfortunately, in cosmology we need more detailed information on solutions. For instance, the inflationary behavior is apparently hidden in some subtle properties of the potential, probably, in complex $\psi$ or $\al$ plane. Moreover, solving the model discussed here is only the first step in looking for observable effects. Thus a sort of classical analytic integrability of homogeneous cosmological models with scalaron is highly desirable. Some relevant results on integrability in static and cosmological models can be found in \cite{ATF1}-\cite{ATF14}.

 \section{Dynamics in $\psi$-version}
 Here we study in detail the $\psi$-version equations, which are important because the potential $v(\psi)$ can often be determined by some field-theoretical model, in which it has a certain `physical' interpretation: the mass squared term of the scalaron, a 'Higgs-type' potential, one of many possible potentials derived in reductions of supergravity (see \cite{Lidsey}-\cite{KL15} and references therein). In connection with inflation, a wider spectrum of potentials was discussed in papers mentioned in this paper. The present author recently discussed a more exotic origin of the scalaron and its potential from affine generalizations of Einstein's gravity theory, \cite{ATF}-\cite{ATFs}.

 \subsection{Main cosmological equations}
 We mostly discuss properties of $\bchi(\psi)$ satisfying equation (\ref{d9}) for $k=0$. Explicit general solutions of this equation can be derived only for very special potentials $v(\psi)$. The simple examples with the exponential or bi-exponential potentials were treated above in the frame of the $t$-version using the explicitly integrable model (\ref{He1})-(\ref{He4}) (see also Appendix~6.2). We can also solve them in the $\psi$-version but the solution is a bit more cumbersome and we'll only give the solution $\bchi(\psi)$ for simple exponential potential. A more complex explicitly solvable model is discussed in Appendix~6.3. For general potentials $v(\psi)$, we derive asymptotic approximations at $\psi \rightarrow \infty$ and  $\psi \rightarrow 0$. For small $\psi$, we also construct power-series expansions.

 The equations in the $\psi$-version immediately follow from (\ref{H1})-(\ref{H2}):
 \be
  \frac{d\al}{d\psi} = \frac{\xi}{\eta}\,, \qquad  2\,\frac{d\eta}{d\psi} \,+\,2(3+c)\,\xi \,+\, \frac{e^{-2c \al}}{\eta} \,\prv(\psi) = 0\,,
 \label{a1}
 \ee
 \be
  6\,\frac{d\xi}{d\psi}\,+\,(3+c)\,\eta\,+\,c \,\frac{e^{-2c \al}}{\eta} \,v(\psi) + (1+c)\,6\,k \,\frac{e^{-2(1 + c)\al}}{\eta} = 0\,,
 \label{a2}
 \ee
 the constraint is unchanged. The gauge-independent equations for $\bxi = \xi e^{c\al},\, \bet = \eta e^{c\al}$ are:
 \be
  2\,\bet^{\pr}(\psi) +\, 6\,\bxi +\, \prv(\psi)\,\bet^{-1} = 0\,, \qquad  2\,\bxi^{\pr}(\psi) +\, \bet + 2\,k\,e^{-2\al}\,\bet^{-1} = 0\,.
 \label{a3}
 \ee
 The gauge-invariant constraint is given in (\ref{Ha}).

 When $k=0$, the constraint and Eqs.(\ref{a3}) give the closed equation for $\bxi(\psi)$:
 \be
  4(\bxi^{\pr}(\psi))^2 \,-\,  6\,\bxi^2(\psi) +\, v(\psi) = 0\,.
 \label{a4}
 \ee
 If we could find $\bxi(\psi)$ we would have the complete solution of the $k=0$ cosmologies. Indeed, $\al^\pr (\psi) = \bchi(\psi) \equiv \bxi/\bet = -\bxi/2\,\bxi^{\pr}$, $\dpsi \equiv \bet(\psi) = -2\bxi^{\pr}(\psi)$,  and it follows that:
 \be
  \al(\psi) = -\int \frac{\bxi(\psi)\,d\psi}{\sqrt{6\,\bxi^2 - v(\psi)}}\,,  \qquad \tau-\tau_0  \,=\,-\int \frac{d\,\psi}   {\sqrt{6\,\bxi^2 \,-\,  v(\psi)}}\,.
 \label{a5}
 \ee

 Equation (\ref{a4}) is essentially equivalent to equation (\ref{d9}) for $\bchi(\psi)$ and, at first sight, it  gives us no new information on discussed problems. However, it is worth of independent study as it contains the potential instead of logarithmic derivative. It can be easily solved in terms of the expansion in powers of $\psi$. To simplify notation we take $x = \sqrt{3/2}\,\psi$, $\,w\equiv\sqrt{6}\,\xi$, $v(\psi) \equiv \tv(x)$ and write the equation and expansions of $v$ and $w$:
 \be
  (w^\pr(x))^2 \,-\,w^2(x) +\, \tv(x) = 0\,, \qquad \tv(x) = \sum_0^\infty v_n x^n\,, \qquad w=\sum_0^\infty  w_n x^n
 \label{a6}
 \ee
 The recurrence relations for $w_n$ cannot be solved in general and are rather cumbersome even for simple potentials, like $\tv = v_0 + v_1 x + v_2 x^2$. The first coefficients for this potentials are:
 \bdm
  w_1 = \sqrt{w_0^2 - v_0}\,,\quad w_2 = 4w_1^{-1}(2w_0 w_1 -v_1) \,, \quad 6 w_3 = w_1^{-1}(w_1^2 - v_2)\,.
 \edm
 We see that this expansion is inconvenient and this also signals that Eq.(\ref{a4}) is not the best starting point for solving even the simplest cosmologies. The perspectives with the large $x$ behavior are even worse. In next Section we find that the $\bchi$-equation is better and derive a good approximation for $x\rightarrow \infty$. With this in mind, let us relate $w(x)$ to $z(x) \equiv \sqrt{6} \bchi(\psi)$:
 \be
  z(x) = w(x)\,[w^2 -v(\psi)]^{-\half}\,, \qquad  w^2(x) =\,\tv(x)\,z^2(x)/(z^2-1) \,.
 \label{a7}
 \ee

 The first equation in (\ref{a7}) is a convenient starting point for \emph{iterations}. We just mention two most evident ideas for this. Taking as a zeroth approximation a physically reasonable $w_0(x)$ we find the zeroth approximation for $z$: $z_0(x)=[1-\tv(x)/w_0^2(x)]^{-\half}$. Then, applying (\ref{d4b}) we can derive the first approximation $w_1(x) = \exp[-2\int z_0^{-1}(x)]$ and hence $z_1$, etc... This procedure is best suited to the case $v<0$ because then $z_n$ is well defined and  $|z_n|<1$.

 We only mention an alternative procedure that looks more complex
 but is applicable to positive potentials. It can be realized by writing the zeroth approximation for $z(x)$ in terms of a reasonable phase portrait for the scalaron, $\eta_0(\psi)\equiv \teta_0(x)$. Indeed, in this case the constraint defines the zeroth approximation for $z_0(x)$ by $z_0^2(x) = 1+\tv(x)\, \teta_0^{-2}(x)$, and the iterations are produced by the inductive relations
 \bdm
  w_n(x) = \exp[-2\int z_{n-1}^{-1}(x)]\,,\quad \teta_n^2(x) = \frac{w_n^2(x)}{z_{n-1}^2(x)}\,, \quad z_n^2(x) = 1+\frac{\tv(x)}{\teta_n^2(x)}\,,\quad n=1,2,3,...
 \edm
 The advantage of these iterations, which are applicable to inflationary scenarios, is that $\eta(\psi)$ describes a phase portrait of the inflaton much discussed in studies of inflation.

 \subsection{Exact and asymptotic solutions of $\bchi(\psi)\,$-\,equation}
 Now we return to equation (\ref{d9}) that looks like a generalized Emden-Fowler equation for a wide class of potentials and is simpler than (\ref{a4}). For some potentials it can be exactly solved and for a rather general potentials in can be solved asymptotically, for $\psi \rightarrow \infty$ and $\psi \rightarrow 0$.
 We rewrite it defining the  `potential' $u(x)\,$ and using the above notation for $\bchi$ and $\psi$:
 \be
  dz/dx = (z^2 - 1) [z\,u(x) + 1] \,, \quad u(x) \equiv d \ln\sqrt{v}/dx\,, \quad z=\sqrt{6}\,\bchi\,,\,\,\, x=\sqrt{3/2}\,\psi\,.
 \label{d10} 
 \ee

 \subsubsection{Solution with $v(\psi)=v_0\,e^{2\,g\psi}$}
 Supposing that $u(x) = g \equiv \tg^{-1} \neq0, \pm1$ we can derive the solution in the form
 \bdm
  2(\tg^{-1} - \tg) (x-x_0) = \ln(\,|z+1|^{1+\tg}\, |z-1|^{1-\tg}\, |z+\tg|^{-2}) \,.
 \edm
 It is easy to find exponentially good approximations for $z(x)$ when z is large or approaching $\,\pm 1$ or $-\tg$. These  can be compared with the corresponding behavior of $\bchi(\psi)$ or $\chi(\al)$ which can be derived for exact solutions given in equations (\ref{He1a})-(\ref{He1b}). That construction used the additional integral $\eta + g\,\xi = C_0$ obtained in the Hamiltonian gauge $c=-3$. It gives the invariant relation between $\bet \equiv \eta e^{-3\al}$ and $\bchi(\psi)$. Then the integral and constraint (\ref{H}),
 \be
  1+g\bchi = C_0\,\bet\,e^{-3\al}\,, \qquad 6\,\bchi(\psi)^2 = 1 + \tv_0\,C_0^{-2}\, (1+g\bchi)^2\, \exp(g\psi+6\al)\,,
 \label{d10a} 
 \ee
 together with relation (\ref{He1b}) define the complete gauge-independent solution of the model with the exponential potential. We can also derive $\chi(\al)$, find the $t$-dependence of $\chi$ and $\bchi$, and thus study on this example the relations between all the three versions.  Such a comparison of versions is also possible with the bi-exponential potential models (\ref{He2})-(\ref{He4}) and (\ref{b1})-(\ref{b7}). For the general potentials, we will find exact solutions in the $\al$-version, while in the $\psi$-version we have only asymptotic solutions to which we now turn our attention.

 \subsubsection{Important transformation of $\bchi(\psi)$ and properties of $v^\pr(\psi)/v(\psi)$}
  A convenient approach to \emph{asymptotic} of $z(x)$ is to introduce the transformation,
 \be
  z(x) = -\frac{1 -\ve e^{-2y}}{1 +\ve e^{-2y}} = -\tanh^{\ve} y(x)\,, \qquad \mathrm{where} \quad \ve =\, \mathrm{sign}(1-z^2) =\, \mathrm{sign}(-v)\,.
 \label{d11} 
 \ee
 Then Eq.(\ref{d10}) becomes simpler and more suited for a qualitative analysis ($\tv(x)\equiv v(\psi)$):
 \be
  dy/dx = 1-\,u(x)\,\tanh^{\ve}(y)\,,\quad\mathrm{where}\quad u(x)=\,\tv^\pr(x)/2\,\tv(x)\,\equiv \tl^\pr(x)\,.
 \label{d12} 
 \ee
 This equation is not difficult to analyze when the logarithmic derivative of the potential has nice properties. Cosmologists often suppose that $\tv(x)$ is a polynomial that does not change sign but vanishes  at $x=0$ like $\tv(x)\sim x^2$. The behavior of the corresponding $u(x)$ at infinity is rather simple. Indeed, consider a polynomial potential, $\tv(x)= P_N(x)$. Then at infinity  $u(x)=(N/2x)[1+\textrm{O}(x^{-m})]$, with integer $m$ satisfying $1\leq m \leq N$. For $u(x)=e^{2gx}P_N(x)\,{\bP}^{-1}_{\bN}(x)$ the asymptotic behavior is $u(x)=g+(N-\bN)\,(2x)^{-1}+\mathrm{O}(x^{-2})$.

 To classify the behavior of $u(x)$ at $x \rightarrow 0$ we suppose that $\tv(x)$ can be represented as a convergent power series $P_{\infty}(x)=\sum v_k\,x^k$:\, 1)~If $v_0\neq 0$, it is easy to show that
 \be
  u(x) = \sum_0 u_m x^m = \frac{1}{2 v_0} (n+1)\,v_{n+1}\,x^n\,[\,1+\sum_1 \,\bu_m x^m\,]\,, \quad \bu_1=\frac{(n+2)v_{n+2}}{(n+1)v_{n+1}}
 \label{d12a} 
 \ee
 for $n\geq 0$, and if $v_k=0$ for $1\leq k \leq n$; 2)~If $v_0 =0$, then for $n\geq 0$, and if $v_k=0$ for $1 \leq k \leq n$, we find the universal behavior for $x\rightarrow 0$:
 \be
  u(x) = \frac{u_\textrm{s}}{x} + \sum_0 u_m x^m = \frac{1}{2 x} (n+1)\,[\,1+\sum_1 \bu_m \,x^m\,]\,,\quad  \bu_1 = \frac{v_{n+2}}{(n+1)\,v_{n+1}}\,.
 \label{d12b} 
 \ee
 The second formula is most general. It is also applicable to singular potentials $\tv(x)$ having terms $\sim x^{-n}$. The simplest behavior for $u(x)$ give pure exponential and pure power potentials, $e^{2gx}$ and $x^n$. All these patterns of asymptotic behavior of the potential $u(x)$ are met in cosmological models and we briefly discuss the most important properties of the solutions for large and small values of $x$ having in mind inflation and other scenarios.

 Some potentials $u(x)$ used in inflationary models (see recent reviews \cite{Encycl}, \cite{Linde-rev}) look, at first sight, different but actually fit into the above classification. For example, the potential derived in \cite{Whitt} for the original Starobinsky model is $v_0(1-e^{-g\psi})^2$, which asymptotically gives $l^\pr(\psi) \simeq -2\,v_0\,g\,e^{-g\psi}$, and $l^\pr(\psi) \simeq -2v_0 /\psi$ for small $\psi$.\footnote{In the additional Section~4.3 we demonstrate that inflation cannot be considered in the domain $\psi \gg 1$ where exist powerful asymptotic expansions in the $\psi$-version. The reason is that the inflationary value of $|z(x)|$ must be large while asymptotically it is close to unity. The adequate approach is the $\al$-version.} The same is true for the more general potentials discussed in \cite{Linde-rev}, $v_0 \tanh^{2n}(g\psi)$. Similar potentials are discussed on connection with `ekpyrotic - bouncing models (see. e.g., \cite{Lehners}, \cite{Battefeld})). A typical potential looks as the following generalization of our bi-exponentials:
 \be
  v(\psi) = v_0(\,e^{g_0\,\psi} + \,v_1\,e^{g_1\,\psi})^{\,m}\,
 \label{d12c} 
 \ee
 where integer $m$ and real $v_0$ are positive (inflationary) or negative (ekpyrotic), $v_1 >0$ if $m<0$, and $g_i$ may have different signs, unlike our model (\ref{He4}).

 \subsubsection{Large $\psi$ behavior of $\bchi(\psi)$}
 Let us first consider $y \rightarrow \pm\infty,\,\,z^2 \rightarrow 1$.  Using that at infinity $\tanh^{\ve}(y)\rightarrow \pm 1$ and supposing that $\tv(x)= \sum  v_n x^n $ at large $x$ is a polynomial of degree $N$ (or, more generally, $e^{gx} \tv(x)$ with $g<6$) we can easily derive the \emph{main terms of asymptotic behavior}:
 \be
  y = x \,\mp\,\ln\sqrt{|\tv(x)|} \,+\, c_0 \,+ \,\mathrm{O}(e^{-2y}) \,,\quad \textrm{for}\,\,\, x \rightarrow  \pm\infty\,,
 \label{d13} 
 \ee
 where $c_0$ is an integration constant and dependence on $\ve$ is hidden in the exponential correction to be derived in a moment. This formula shows that, for polynomial potentials, $y\sim x$ and we can obtain the main terms of the asymptotic expansions of $y$, $z$ for $x\rightarrow +\infty$:
 \be
  \sqrt{6}\,{\frac{d\al}{d\psi}} = \sqrt{6}\,\bchi(\psi)\,\equiv \,z(x) = -\tanh^{\ve}\,[\,x \,-\,\ln\sqrt{|\tv(x)|} \,+\, c_0\,] \,+ \,\mathrm{O}(e^{-4x}) \,.
 \label{d14} 
 \ee
 Using this and higher asymptotic approximations we can find the corresponding \emph{asymptotic expansions for the portrait} $\al(\psi)$ and, by inversion, $\psi(\al)$. In the simplest approximation
 \be
  \sqrt{6}\,{\frac{d\al}{d\psi}}=-[1-2e^{-2(x+c_0)}\,\tv(x)+...]\equiv  -[1-2C_0\,v(\psi)\exp(-\sqrt{6}\,\psi)+...]\,,
 \label{d14a} 
 \ee
 where $C_0$ is the redefined arbitrary constant. By integration we can derive the asymptotic portrait $\al(\psi)$. Taking the popular `cosmological' potential $v(\psi)= v_0\,\psi^2$ we find
 \bdm
  \sqrt{6}\,\al(\psi) = \psi_0-\psi + \bC_0\,[1-(3\,\psi^2+\sqrt{6}\,\psi+1)\exp(-\sqrt{6}\,\psi)+...]\,,
 \edm
 where $\psi_0$ is the integration constant. The inverted expression defines the asymptotic of $\psi(\al)$.

 Note that this approach to the asymptotic behavior at infinity in fact uses \emph{iterations} with the zeroth approximation given by (\ref{d13}). To find the next one we write the exact equation,
 \be
  y(x) = y_0+\int u(x)\,[1-\tanh^\ve y(x)]\,=\,y_0+c_0-2\int_x^\infty dx\,u(x)\,e^{-2y}\,(1+\ve e^{-2y})^{-1}\,.
 \label{d14b} 
 \ee
 where $u(x)=\tl^\pr(x)$ and $y_0(x)\equiv x - \tl(x)$ is the zeroth approximation. Replacing $y$ by $y_0$ in the r.h.s. we obtain the first approximation $y_1$, in which we may expand the integrand in powers of $e^{-2y_0}$ and find the first exponentially small correction to $y_0$ also independent of $\ve$:
 \be
  y_1(x) = x -\tl(x) + c_0-2\int_x^\infty \tl^\pr(x)\,e^{\,2\,\tl(x)-2x\,}\,+\,\mathrm{O}(e^{-4y_0})\,.
 \label{d14c} 
 \ee
 Remembering the above definition, $\tl(x) = \ln{\sqrt{\tv(x)}}$, we find that in this approximation
 \be
  y_1 = x-\sqrt{\ln \tv}\,+\,c_0-\int_x^\infty \frac{d\tv}{dx}\,e^{-2x\,}\,=\,\sqrt{\frac{3}{2}}\,\psi - \sqrt{\ln v}\, + c_0-\int_\psi^\infty \frac{dv}{d\psi}\,e^{-\sqrt{6}\,\psi},
 \label{d14d}
 \ee
 where the integral can be expressed in terms of elementary and special functions for a wide class of the potentials. The new approximation $\sqrt{6}\,\psi = \tanh^{\ve} y_1$
 is \emph{significantly better} than (\ref{d14}) and can be further improved if necessary.

 \subsubsection{Small $\psi$ behavior of $\bchi(\psi)$}
 The behavior of $z(x)$ and $y(x)$ near $x=0$ is more complicated because $u(x)$ for $x \rightarrow 0$ may be zero, constant or infinite. In addition, when $\ve=-1$, the solutions $y(x)$ must not vanish for $x \rightarrow 0$ if $u(x)$ is singular or constant at $x=0$. For this reason we first consider the case $\ve=+1$. Then for the regular positive potentials that vanish for $x \rightarrow 0$ it is easy to derive a universal approximation that can be used for subsequent iterations\footnote{This is a special solution of the linearized equation (\ref{d12}), the general one to be discussed in a moment.}
 \be
  y_0(x) = [\tv(x)]^{-1/2}\int_0^x{[\tv(x)]^{1/2}} + ...\,,\quad \textrm{if}\,\,\,\,  y_0^2 \ll 1\,; \qquad y^\pr_0 = 1 - u(x)\,y_0\,.
 \label{d15} 
 \ee
 This solution is \emph{independent} on any arbitrary parameter. This means that it may be either an enveloping solution or a separatrix (if it has no common points with any other solution).

 Instead of using iterations one can directly construct a \emph{power-series expansion}. However, for qualitative analysis of solutions with realistic potentials the general approach of (\ref{d15}) may prove preferable. An important example is the \emph{ parameter-independent} solution for potentials (\ref{d12b}).
 When $u_\textrm{s}\neq 0$, the solution of (\ref{d12}) with $\ve=+1$ must vanish when $x\rightarrow 0$. Expanding it in the series $y(x) = \sum_{n=1} a_n x^n$ we can find all $a_n$ with $n > 1$ in terms of $a_1$:
 \be
  y(x) = \frac{x}{1+u_\textrm{s}}\,\bigg[1 - \frac{x u_0}{2+u_\textrm{s}}\,-\,\frac{x^2}{3+u_\textrm{s}} \bigg( u_1 - \frac{u_0^2}{2+u_\textrm{s}} - \frac{u_\textrm{s}}{3(1+u_\textrm{s})^2}\bigg)
  + \mathrm{O}(x^3)\bigg] \,.
 \label{d15a} 
 \ee
 Here we take into account the three coefficients $a_i$ and the third term in $a_3$ is the contribution of the $y^3$. This formula and its extensions to higher order terms are applicable to arbitrary parameters. In particular, if all $u_n=0$ and $u_\textrm{s}=\sigma$ is any real number, it gives $y(x)$ for the pure power potential $v=x^{2\sigma}$. The two first terms in (\ref{d15a}) agree with corresponding Eq.(\ref{d15}) and both methods can be generalized to derive the one-parameter family of solutions near $x=0$. They are based on splitting $y(x)$ into `big' and `small' parts, $y \equiv y_0(x)+y_p(x)$, so that $y_p$ could either be  expanded in a regular power series, like (\ref{d15a}),  or satisfy a soluble (e.g, linear) differential equation, like that one defining $y_0$ in (\ref{d15}).

 In deriving solution (\ref{d15}) we suppose that $y_0=0$ and then find the \emph{unique small solution} $y_1$ for arbitrary potentials $u(x)$. If the $u$-potential is regular, $u_\textrm{s}=0$, we can construct the general solution $y(x) = y_0(x) + y_1(x) = a_0 + a_1 x + a_2 x^2 +...$ by taking  $y_0=a_0$ and applying the addition theorem to $\tanh(a_0+y_1)$.  Then we either use for $y(x)$ the linear approximation in $y_1$ or directly expand $y(x)$ in the power series in $x$. We first write the \emph{general expansion} for $\tanh^\ve(y_0+y_1)$ in powers of $y_1$, with notation $t_0\equiv \tanh(y_0)$ and $t_1\equiv \tanh(y_1)$:
 \be
  \tanh^\ve y(x) = \tanh^\ve(y_0+y_1) = t^\ve_0 + (1-t_0^{2\ve})\,\,t_1\,(1\,+\,t_1\,t^\ve_0\,)^{-1}\,.
 \label{d15b} 
 \ee
 Applying this to constant $y_0 = a_0$ and approximating $t_1 = y_1 + \textrm{O}(y_1^3)$ we can easily find the linear equation for $y_1$, the solutions of which generalizes (\ref{d15}). \emph{Alternatively}, we may use (\ref{d15b}) to derive the expansion in powers of $x$. With arbitrary $a_0$ and $a_1 = (1-u_0 t_0), $ we find
 \be
  y(x)= a_0 + a_1 x - \half [\,u_1 t_0 + u_0\,c_0^{-2}a_1]\,x^2 -\third \{\,u_2 t_0 + c_0^{-2}[u_1 a_1 + u_0(\,a_2 - t_0a_1^2\,)]\}x^3- \textrm{O}(x^4)\,,
 \label{d15c} 
 \ee
 where $t_0 \equiv \tanh(a_0)$, $c_0 \equiv \cosh(a_0)$, and $a_2$ is the $x^2$-coefficient in this expansion. The expression for $z(x)$ can be written with the help of the same formula (\ref{d15b}). Depending on the concrete values of the parameters it may sometimes be used for extrapolation of $z(x)$ to asymptotic regions. It also is useful in comparing the small $x$ behavior of $z(x)$ for different potentials. Note that this solution can easily be rewritten for $\ve=-1$.

 The elementary formula (\ref{d15b}) is extremely useful for deriving various approximations. For instance, we can take as $y_0$ asymptotic approximation (\ref{d13}) and find $y_1$ by solving the linear equation obtained from (\ref{d12}) by linearizing $\tanh^\ve(y_0 + y_1)$ in $y_1$. Then we derive $z(x)=-\tanh^\ve(y_0 + y_1)$ using Eq.(\ref{d15b}). This equation (\ref{d15b}) may find most interesting applications in studies of the small $x$ behavior of $z(x)$, especially for singular $u$-potentials and for the $\ve=-1$ case, when the behavior of the solutions at the origin $x=0$ is more complex.

 In fact, to \emph{compensate the singularity} $u_\textrm{s}/x$, one must take $y_0 = a\ln x + a_0$, where $a = u_\textrm{s}$. Then, supposing that $y_1= \sum_1^{\infty} a_n x^n$, $a_0\equiv \ln \ba$ and thus $y_0=\ln (\ba x^{a})$, we can apply (\ref{d15b}) to solve Eq.(\ref{d12}) for both signs of $\ve$. Indeed, defining $a_0\equiv \ln \ba$, we immediately see that
 \be
  \tanh^\ve(y_0 + y_1) = -(1+\ve\ba^2 x^{2a})^{-2}\,[1-\ba^4 x^{4a} - 4\ve \ba^2 x^{2a}\,(y_1 + \textrm{O}(y_1^2))]\,.
 \label{d16} 
 \ee
 With this relation we can either expand $y_1(x)$ in the power series and solve several recurrence relations or, instead, solve the linearized equation for it. We only illustrate the first approach by writing a few terms for the simplest singular potential $u(x) = u_\textrm{s}/x$:
 \be
  y(x) = \ln(\ba x^2) + x -\ve \ba^2(1+4x/3)\,x^2 +...
 \label{d17} 
 \ee

 In the beginning of this Subsection we met only one analytically solved equation, on which one can roughly test the proposed approximation methods. The simpler explicitly solved example in the $\psi$-frame is presented in Appendix~6.3. On these examples one can  check the precision of the approximate solution derived here in more detail. In this Section, we briefly described results of our studies in the general $\psi$-version. This in principle would allow us to find, in main models of cosmology, the approximate $(\al,\psi)$ portraits which are exact in the asymptotic regions. A more complete presentation would be possible if we concentrated on specific inflationary or ekpyrotic-bouncing models not contradicting to the present observational data. This task requires quite different means and exceeds the author's space-time capabilities, but in next Section we derive exact solutions that allow us to find a very simple picture of inflation. In addition, this will help us to better understand and derive inflationary solutions $y(x)$. Also note that the general `$\al$-solutions' can be used in analyzing some ekpyrotic-bouncing cosmological scenarios.

 \section{Dynamics in $\al\,$-version}
 Here we present an unusual \emph{general approach to generating cosmological solutions}. It is based on simple examples of relations between the solutions in the $\psi$-version and the $\al$-version (see (\ref{Hc2}), (\ref{H2e}), (\ref{c1})-(\ref{c7})). The close mathematical connection between the two versions is given by the functions $\bchi(\psi)$ and $\chi(\al)$ introduced and discussed above. Now we first derive the exact analytic solution of equation (\ref{d5}) for $\chi^2(\al)$ with arbitrary $\bl(\al)$. Then we write  the exact analytic solutions of all dynamical equations, formulate a general approach to inflationary phenomena and compare it to the standard one.

 \subsection{Exact solution of $\chi^2(\al)\,$-equation for $k=0$}
 We start by solving the $k=0$ equation because to solve the general one,
 \be
  \frac{d\chi^2}{d\al} = \bigl(\chi^2 - 6\bigr) \bigl(\chi^2 + \bl^{\pr}(\al)  \bigr) + \frac{2k}{\xi^2(\al)} \,e^{-2(1+c)\,\al} \bigl(\chi^2 + 3 \,\bl^{\prime}(\al)  \bigl) \,,
 \label{d5c}
 \ee
 one has to know $\xi^2(\al)$. When $k=0$, we the obvious general solution is simply
 \be
  \chi^2 (\al) = 6 - e^{6\al} \,\bv(\al) \bigl[\,C_0 \,+ \int e^{6\al}\,\bv(\al)\,\bigr]^{-1}\,, \quad k = 0 \,,
 \label{d6}
 \ee
 where $C_0$ is an arbitrary constant.  This solution can be verified by inserting $\chi^2 (\al)$ into equation (\ref{d5}) and taking account of the above definition $\bl (\al) \equiv \ln \bv(\al)$. The main problem with this solution is to find the conditions for positivity of the r.h.s. of Eq.(\ref{d6}). We discuss it on the particular example of the exponential $\al$-potential, $\bv(\al) = v_0 \exp (g\al)$,
 \be
  \chi^2\,(\al) = 6 - (g+6)\,[\,1 +\, C_1\,e^{-(6+g)\,\al}\,]^{-1}\,, \qquad C_1 \equiv C_0\,(g+6) \,v_0^{-1} \,,
 \label{d7}
 \ee
 which is positive for all $\al\,\ep\,\Re$ if and only if $\,C_1 > 0, \,g < 0,\,g\neq -6$. Otherwise, it is not difficult to find solutions becoming negative on some intervals belonging to $\,-\infty < \al < +\infty\,$. In a moment we consider the case $C_1<0,\, g<0$ when the solution is rather complex.

 To simplify discussions of these complex solutions let us rewrite $\chi^2$ in $N/D$ form,\footnote{Note that the main properties of $\chi^2$ are defined by the signs of $g$ and $C_1$, with the proper account of the restriction on the sign of $(\chi^2-6)$ according to equation (\ref{d5b}).}
 \be
  \chi^2 \,= -g(\, \tau - \,C_2\,) (\, \tau +\, C_1\,)^{-1}\,; \qquad C_2 = 6\,C_1/g\,,\quad \bg \equiv (g+6)\,, \quad \tau \equiv e^{\,\bg \,\al}\,.
 \label{d8}
 \ee
 Then it is clear that for $\,C_1 > 0, \,g > 0$ the \emph{nominator has one zero} at $ \tau = -C_2$ and becomes negative for $ \tau > |C_2|$, where  $\al > \al_0 \equiv \bg^{-1} \ln |C_2|$. This means that $\al_0$ is the \emph{branch point in the complex $\al$-plane} and there exists the \emph{second sheet of the Riemannian surface} of the analytic function $\chi(\al) \equiv \sqrt {\chi^2(\al)}$ having the `physical' cut $-\infty < \al \leq \al_0\,$, with $\chi(\al_0) = 0$; on the upper edge of the cut $\chi(\al)$ is positive and on the lower edge it is negative. To find $\psi(\al)$ one should integrate $\chi(\al)$ along the cut; $\al_0$ is the extremum of $\psi(\al)$.

 The singularities of $\chi(\al)$ corresponding to \emph{zeroes of the denominator} are stronger because then $\chi^2(\al) \rightarrow \infty$, but they are also integrable. The simplest such case is $\,C_1 < 0, \,g > 0$, when $\chi^2 < 0$ for $ \tau > |C_1|$ and is infinite at $ \tau = |C_1| \equiv e^{\bg\,\al_1}$, where $\chi^2 \sim \bg\,(\al_1 -\al)^{-1}$. The same behavior has the \emph{special exact solution} that can be obtained from Eq.(\ref{d7}) in the limit $\bg \equiv (6+g) \rightarrow 0$: if we take $|C_1| = 1 + \bg\,\al_0\,$ we find that the limit is $\chi^2\,(\al) = 6 + (\al_0 - \al)^{-1}$ (this solution can also be derived from Eq.(\ref{d5}) for $k=0$, $\bl^{\pr}(\al) = -6$).

 A configuration with more singularities emerges in the case $\,C_1 < 0, \,g < 0$. Then the `positive support' of the solution (\ref{d8}), where  $\chi^2$ is positive, consists of two separate intervals:
 \bdm
  0 < \tau < |C_1|,\,\,\, C_2 < \tau < \infty\,,\,\mathrm{if} \,\,\bg > 0; \quad 0 < \tau < |C_2|,\,\,\,|C_1| < \tau < \infty\,, \, \mathrm{if}\,\, \bg < 0\,.
 \edm
 One can see that there is a fundamental difference between solutions with different number of branch points, which also depends on their behavior (singular or regular). In fact, there exist five types of the solutions: R (no b.p.), 1R (1 regular b.p), 1S (1 singular b.p.), 2SR (singular and regular b.p.). The support of the solutions 1S and 1R does not include large enough values of $\al$ and thus for them the metric has a finite upper bound. We may call them \emph{gravitationally regular solutions}. All other solutions do not have this property.

 Now it is possible to derive $\psi(\al)$ by integrating $\prpsi (\al) \equiv \chi(\al)$ along the `physical' paths in the complex $\al$-plane and thus to find the portraits of cosmologies with the simple exponential potential and $k=0$. The same consideration can be applied to analyzing solutions for other potentials, e.g., $\bv(\al)=e^{\,g\al}\, P_n(\al)$. This class is rich enough to describe different scenarios, like inflation, bouncing, etc.

 \subsection{Exact solutions $\eta^2(\al)$, $\xi^2(\al)$, $\chi^2(\al)$ for arbitrary $\bv(\al)$ and $k$}
 Here we show that Eqs.(\ref{H1})-(\ref{H2c}) can be transformed into linear equations for functions $\eta^2(\al)$, $\xi^2(\al)$, which can be integrated for arbitrary preassigned potential $\tv(\al)$ and with  arbitrary curvature parameter $k$. It is more convenient to consider equations (\ref{H2}) and (\ref{H2c}), which we transform to the $\al$-picture using (\ref{d2}) and other above definitions. Thus we derive two linear differential equations equations that can be solved with any given $\bv(\al)$:
 \be
  \frac{d\eta^2}{d\al}  \,+\, 2(3+c)\,\eta^2 \,+\, e^{-2c \al}\, \bv^{\prime}(\al)\,= 0\,, \qquad  \frac{d\xi^2}{d\al}\,+\,2c \,\xi^2 \,+\, \eta^2 \,+\,  2k\,e^{-2(1 + c)\al} = 0\,.
 \label{sq1}
 \ee
 Now, introducing new positive functions $y(\al) = e^{2c \al} \,\eta^2$ and $x(\al) = e^{2c\al} \,\xi^2$, we find for $y(\al)$ and $x(\al)$ the gauge-independent equations and constraint
 \be
  \pry(\al)\,+\, 6y(\al) \,+\, \bv^{\prime}(\al)\,= 0\,, \qquad \prx(\al)  \,+\, y(\al) \,+\,  2k\,e^{-2\al} = 0\,.
 \label{sq2}
 \ee
 \be
  y(\al) - 6 \,x(\al) \,+\, \bv(\al) + 6k \,e^{-2\al} \, = \,0\,.
 \label{sq3}
 \ee

 Keeping in mind Eq.(\ref{d6}) we write the solution of the first equation (\ref{sq2}) in the form,
 \be
  y(\al) = 6 \,e^{-6\al} I(\al) \,-\, \bv(\al) \,; \qquad I(\al) \equiv [\,C_0 + \int_{\al_{-}}^{\al} e^{6\al}\, \bv(\al)\,]\,,
 \label{sq4}
 \ee
 where $C_0$ and $\al_{-}\geq -\infty$ are real numbers. Then we find $x(\al)$ from constraint (\ref{sq3}):
 \be
  x(\al) = e^{-6\al} I(\al) \,+\, k \,e^{-2\al} \,.
 \label{sq5}
 \ee
 It can also be derived from (\ref{sq2}) but then the additional arbitrary constant must be fixed by constraint (\ref{sq3}), which shows that the integrals $I(\al)$ in (\ref{sq4}) and (\ref{sq5}) are identical. Now we can find the (gauge dependent) phase portrait of cosmology in the  $\al$-version,
 \be
  \xi = \dal = e^{-c\al}\, \sqrt{x(\al)} = e^{-(1+c)\al}\,[\,e^{-4\al} I(\al) \,+\, k\,]^{1/2} \,,
 \label{sq6}
 \ee
 which gives $t(\al)$ by one integration over `physical' cuts. As an exercise, one may derive explicit expressions for $\al(t)$ for the potentials $v = e^{g\al} P_n(\al)$.

 The obtained solutions also give the \emph{exact} expression for $\chi^2(\al)$ in the case $k \neq 0$:
 \be
  \chi^2(\al) \equiv y(\al)/ x(\al) = [\,6 \,e^{-6\al} I(\al) \,-\, \bv(\al)\,]\,[\,e^{-6\al} I(\al) \,+\, k \,e^{-2\al}\,]^{-1} \,,
  \label{sq7}
 \ee
 which, of course, coincides with (\ref{d6}) when $k=0$. It is not difficult to check that this expression satisfies complete differential equation (\ref{d5}) if we take into account (\ref{sq2}), (\ref{sq3}). Moreover, if we substitute into equation (\ref{d5}) the expression for $\xi^2 = e^{-2c\al} x(\al)$ from (\ref{sq5}), we obtain the well-defined equation for $\chi^2(\al)$, but it is much simpler to use the exact solutions.  Anyway, the approach to analysis of the singularities and support of the most general solution is essentially the same as above -- we first look for zeroes of the nominator and denominator and then find the cuts, on the edges of which we should integrate $\chi(\al) \equiv \prpsi(\al)$ to find $\psi(\al)$. In principle, this result could allow us to establish a correspondence between the two versions. In practice, this is a not so simple task, which requires a careful investigation.

 At the moment, the best strategy is to compare physical results for interesting classes of potentials in the new version with those obtained for the well studied potentials of the standard $\psi$-version.  A more radical approach is to try to find, directly in the new version, potentials that describe physically interesting phenomena, like inflation, bouncing, or something else. As we demonstrated above, this approach can be significantly strengthened by using in addition other inputs -- various portraits motivated and supported both  by theoretical intuitive ideas and observational data. The synergetic strategy of using the full mathematical structure outlined above -- all gauges, versions, and inputs -- looks like a promising global approach to isotropic cosmology that possibly could help us to return, sooner or later, in the higher-dimensional world of real physics.

 \subsubsection{On replacing potential by kinetic energies}
 It is usually supposed that the dynamical functions $\xi$, $\eta$ can be expressed in terms of the scalaron potential $v(\psi)$, which is unknown and is usually chosen to satisfy some `reasonable' properties providing a sort of inflation or other phenomena. It is supposed that the potential can be somehow derived in a future superstring theory or in present supergravity considerations. Our $\al$-version may suggest a different approach to finding the potential -- first guessing the scalaron kinetic `energy' $\eta(\al)$ and then deriving $v(\al)$ and $\xi(\al)$ as simple functionals of $\eta$. To derive the exact expression we simply integrate the \emph{relations} (\ref{sq2}) (forgetting that they are the differential equations for $x(\al)$, $y(\al)$) and find:
 \be
  \bv(\al) = -y(\al) + 6J(\al)\,, \quad  x(\al) = k e^{-2\al} + J(\al)\,; \qquad J(\al) \equiv C_1+\int_\al^{\al_{+}} y(\al) \,.
 \label{sq2a}
 \ee
 Here we defined $J(\al)$ similar to $I(\al)$ introduced in (\ref{sq4}) and with the same aim -- to simplify considering the positivity conditions and asymptotic behavior for $y(\al),\, x(\al),\, \bv(\al)$.

 By the way, the expression for the fundamental function $\chi(\al)$ in terms of $y(\al)$,
 \be
  \chi^2(\al) = y\,[ke^{-2\al} + J(\al)]^{-1} =\,
  -L^\pr(\al)\,[\,k e^{-L(\al) - 2 \al}+1\,]^{-1}\,, \qquad L(\al) \equiv \ln{J(\al)}\,.
 \label{sq2c}
 \ee
 is simple and physically transparent. It demonstrates a deep connection between the scalaron kinetic energy and metric and permits to reconstruct all the portraits of our cosmology. The portrait $\dal(\al)$ is explicitly given by (\ref{sq2a}); the portrait $\psi(\al) = \int \chi(\al)$ can be derived from (\ref{sq2c}) (for arbitrary $k$). As far as we can derive the inverse function $\al(\psi)$ we also find $\dpsi(\psi)$ and $v(\psi) \equiv \bv[\al(\psi)]$. Note also that $\chi=-dL/d\psi$ for $k=0$ to compare with Eqs.(\ref{i6}),(\ref{i9}).

 One final remark on relation of this construction to inflationary models. The creators of inflation observed that very different potentials may give very similar inflationary scenarios. In fact, in the early Linde models the main idea was to find potentials that define inflationary `motions' of the inflaton. The basic ingredient was these specific motions and the potential was an instrument to describe the model in a more standard field theoretic frame. Possibly, the expressions of the potential in terms of the kinetic energy of the scalaron depending on the metric may give a different, mathematically accurate realization of these ideas.

 Another option is to take as an input the main dynamical characteristic of the Universe, the Hubble parameter $H^2(\al)\equiv x(\al)$, together with  $\chi(\al)$ determined by Eqs.(\ref{d4c}), which is even simpler to use than Eq.(\ref{sq2c}). This requires some work and time for customization but may give a new insight into the structure of cosmological models. One may call such an approach `constructive' cosmology having in mind that the mathematical structure of the classical isotropic cosmology is an instrument transforming the input portraits or potentials into particular cosmological scenarios to be eventually confronted to the  observational data. This idea looks not very appealing to theorists although some hints to `reconstructing' the potential $v(\psi)$ from observational data, can be traced in literature.~\footnote{After finishing this paper I became aware of Ref.~\cite{Star98} proposing to determine $v(\psi)$ from some contemporary observational data on FLRW cosmology and, as a matter of fact, using for this purpose an $\al$-version-type equation for the Hubble parameter.}

 \subsection{A fresh look at inflation and inflationary perturbation theory}
 The standard approach to inflationary models does not contradict to the recent observational data, even if one uses various potentials and considers non-minimal inflaton coupling to gravity. However, it is not a completed theory as it actually describes the short period of the classical expansion and the emergence of quantum fluctuations on the classical background.  At the moment, there exist a few other models also explaining main properties of the very early Universe but giving predictions that can in future be distinguished from those of inflationary models.\footnote{See, e.g., a discussion of some quantum aspects of the Universe creation in \cite{Lehners} that is related to a modernized version of the classical ekpyrotic-plus-bouncing dynamics. Our approach is applicable also to these more complex models. For a review of this an other alternatives to inflationary models see \cite{Battefeld}.} 
 Our aim in this  Section is to apply our analysis of global properties of the cosmological solutions and, especially, of the exact $\al$-version solutions to identifying classical initial conditions and their consistency with the inflationary restrictions which we discuss in a moment. In this way, we hope to clarify the meaning of some specific features of the inflationary models and thus to make their comparison to other approaches easier.

 We begin with the standard conditions for inflation and the parameters accessible to measurements. In our language, the obvious \emph{necessary condition for inflation} is $|\bchi(\psi)| \gg 1$ on a small interval of $\psi$, or, equivalently, $\chi^2(\al) \ll 1$ on the corresponding large interval $(\al_{i}< \al < \al_{f})$, where  $\al_{f}-\al_{i} = N_e \sim 50$ is the so-called number of $e$-folds  (see, e.g., \cite{Rubakov}, \cite{Linde-rev}). Using equations (\ref{H2a}) and (\ref{sq2}) with $k=0$\, we derive the \emph{exact relations}\footnote{The standard inflationary models suppose that $k=0$, and in this Section we always keep this condition.}
 \be
  -2\,\dxi/\xi^2 \equiv -\prx(\al) /x(\al) = y(\al)/x(\al) \equiv \chi^2(\al) \equiv 2\,\hep \equiv 1/\bchi^2(\psi)\,.
 \label{i1}
 \ee
 Here we define the new function $\hep$ of one of the three variables ($\tau$, $\al $, $\psi$), which is supposed to be small, $\hep \ll 1$, and approximately constant on the corresponding inflationary intervals $(\tau_{i},\tau_{f})$, $(\al_{i},\al_{f})$,  $(\psi_{f},\psi_{i})$. Under these conditions, we call it the first inflationary parameter that is usually denoted by simple $\ep$. Applying now the exact constraint (\ref{sq3}) we derive
 \be
  \hr \equiv \dpsi^2 /v(\psi) \equiv y(\al)/\bv(\al) = \frac{\hep}{3} \,(1-\hep/3)^{-1} \simeq \frac{\hep}{3} \,,
 \label{i2}
 \ee
 where the last approximation (important for the standard considerations) is valid if $\hep \ll  3$ ($\chi^2 \ll 6$).\footnote{In view of the factors $(1/6)^n$, which are most clearly seen in equation (\ref{i4}) below, the inflationary conditions can actually be somewhat weaker, like  $\hep \leq 1/2\,$. The exact bound depends on the potential.}  The approximation can be applied only to positive potentials, and positivity is necessary at least in the inflationary domain. If $\chi^2 \geq 1$, the inflation stops and, for $\chi^2 \rightarrow 6$, the scalaron kinetic energy dominates over the potential one. If  $v>0$, all the solutions must satisfy the condition $\chi^2 \equiv 1/\bchi^2 < 6\,$; they can be only asymptotically close to the limiting curves $\chi =\pm \sqrt{6}\,$ ($\bchi = \pm 1/\sqrt{6}\,$). Near these attractors the parameter $\hr$ becomes infinite and the scalaron potential energy is negligible. As demonstrated in Section~3, this behavior of $\bchi$ is possible for large and small $\psi$. In contrast, the condition $|\bchi|\gg 1$ is generally more difficult to satisfy in the $\psi$-version. It requires $\ve =-1$ in (\ref{d11}) and is  possible for small enough $\psi$, because $\sqrt{6}\bchi(\psi) \equiv z = \coth y\,$ can be large for small $y$ only (see (\ref{d10}), (\ref{d12})). Here, we do not attempt to investigate this part of the $\psi$-picture depending on unknown concrete potentials and initial conditions but describe an easier and more rigorous approach to inflation.

 Starting with the model-independent definition of inflation given above in (\ref{i1})-(\ref{i2}) we now show that the $\al$-version gives the detailed picture of inflationary cosmology. In the standard approach to inflation,  cosmologists use one more parameter $\het$ defined by
 \be
  -2\,\frac{\deta}{\eta \xi} \equiv -\frac{\pry(\al)}{y (\al)} \equiv 2(\het - \hep)\,, \qquad \het - \chi^2(\al) = -\frac{\chi^\pr(\al)}{\chi(\al)} \equiv -\frac{d\chi}{d\psi} \equiv \frac{\bchi^\pr(\psi)}{\bchi^2(\psi)}\,,
 \label{i3}
 \ee
 where the second formula gives \emph{the exact expression} of $\het$ in terms of $\chi(\al)$ or $\bchi(\psi)=1/\chi$. The inflationary parameters can be derived in terms of $\bv(\al)$ by using solutions (\ref{sq2})-(\ref{sq5}) and we can expand them in the derivatives $\bv^{(n)}(\al)$ applying the simple formula ($k=0$):
 \be
  6 x(\al) \equiv 6 e^{-6\al} I(\al)= C_0 e^{-6\al} + \sum_{0}^{N} (-6)^{-n} \bv^{(n)}(\al) - (-6)^{-N} e^{-6\al} \int_{-\infty}^\al e^{6\al}\,\bv^{(n+1)}(\al)\,,
 \label{i4}
 \ee
 where $C_0$ is an arbitrary integration constant defining the unique solution $[x(\al;C_0), y(\al;C_0)]$, $y(\al)$ is given by (\ref{sq2}). The lower integration limit is chosen supposing the effective  potential $e^{6\al}\,\bv(\al)$ vanishes in the extreme `quantum' limit $\al \rightarrow -\infty$. A very useful property of the solution for the chaotic inflation is that the ratio of the first term to the rest is exponentially small. This means that in the `classical' domain $\al > \al_i \gg 1$ we may approximately use the solution with $C_0=0$. Neglecting this term in the solution $[x(\al), y(\al)]$, we find
 \be
  \hr(\al) \equiv \,\frac{y(\al)}{\bv(\al)}\,=\,
  6\,\frac{x(\al)}{\bv(\al)}- 1 \,=\, \frac{\chi^2}{6}\, \biggl(1-\frac{\chi^2}{6}\biggr)^{-1}=\,\, \sum_1^{\infty} (-1)^n \frac{\bv^{(n)}(\al)}{6^n\,\bv(\al)} \,.
 \label{i5}
 \ee
 The properties of the series are defined by the potential $\bv(\al)$ and, in principle, can be derived from the exact analytical solution given by equations (\ref{sq4}), (\ref{sq7}).\footnote{To avoid worries on the series convergence, we may have in mind polynomial potentials $\bv(\al)$ without zeroes in the domain of our interest, and thus the series has a finite number of terms.} Thus we suppose the series defines a regular function $\chi^2$ that is small in some interval $1< \al_i<\al<\al_f$. In addition, we assume it is sufficient to take into account just a few terms of the expansion, which give the corrections to standard inflationary parameters. In this sense, Eq.(\ref{i5}) can be considered as a somewhat unusual perturbative expansion. To relate it to the standard inflationary language we \emph{translate} (\ref{i5}) into the $\psi$-version by a \emph{perturbative construction} of $\chi$.

 The lowest approximation for $\chi$ is given by the first term of the sum in (\ref{i5}):
 \be
  \chi^2 \cong -\frac{\bv^\pr(\al)}{\bv(\al)} \equiv -\bl^\pr(\al) \equiv -\chi \,l^\pr(\psi) \equiv  -\chi\, \frac{v^\pr(\psi)}{v(\psi)}\,,\quad \chi \cong \chi_1 = -l^\pr(\psi)\,;
  \quad \het_1 \cong \frac{v^{\prime\pr}(\psi)}{v(\psi)}\,,
 \label{i6}
 \ee
 where $\het_1$ is derived from (\ref{i3}) with $\chi$ approximated by $\chi_1$ and, in the same approximation, $2\hep \cong \chi_1^2 = (l^\pr)^2$. Here we neglect the terms with $n\geq 2$, which can be calculated recursively and are sums of the monomials $\prod_{i=1}^n(\bl^{(i)})^{\,k_i}$ with $\sum ik_i =n\,$, as can be seen from relations:
 \be
  \frac{\bv^{(n+1)}}{\bv(\al)}\,=\,\frac{d}{d\al}\frac{\bv^n(\al)}{\bv(\al)} +\, \bl^\pr(\al)\frac{\bv^n(\al)}{\bv(\al)}\,;\qquad
  \frac{\bv^\pr(\al)}{\bv(\al)}\,=\,\bl^\pr(\al)\,, \quad
  \frac{\bv^{\pr\pr}}{\bv}\,=\,\bl^{\pr\pr} +\,(\bl^\pr)^2\,,...\,;
 \label{i7}
 \ee
 we need not write the same expressions for $v^{(n)}(\psi)/\,v(\psi)$ in terms of $l^{(n)}(\psi) \equiv l^{(n)}$. It must be emphasized that the `logarithmic' representations for $\chi^2$ and $\chi$ (invariant under \emph{$v$-scaling}, i.e., $v\mapsto c_0 v$) are strictly valid only for solutions with $C_0=0$ in (\ref{i4}).\footnote{The general solution has the non-scaling  non-logarithmic term  $C_0\,e^{-6\al}/\bv(\al)$ in the (\ref{i5}).} We have $v$-\emph{scale invariance if and only if} $C_0=k=0$. This is a \emph{most general definition of inflationary} $\chi\,$.

 Now let us find a few first terms for the expansion of $\chi$, which give corrections to the standard inflationary parameters. Defining $d_\al \equiv d/d\al,\,\, d_\psi \equiv d/d\psi\,$ and recalling that $d_\al \equiv \chi d_\psi\,$, we express the operators $d^n_\al$ in terms of powers of $d_\psi$ and of $\chi$:
 \be
  d_\al^2 = \chi[\,\chi d_\psi^2 +(d_\psi\chi)d_\psi]\,\,,\,\,\, d_\al^3 = \chi[\,\chi^2 d_\psi^3 + 3\chi (d_\psi\chi)\,d_\psi^2 + d_\psi(\chi d_\psi\chi)\,d_\psi],...\,\,\,d_\al^n = \chi D_n\,,
 \label{i8}
 \ee
 where the operator $\chi D_n$ can be calculated recursively. This operator is homogenous and of the same order $n$ in both the variables $d_\psi$ and $\chi$. This means that the $n$-th term of the perturbative expansion (\ref{i5}) is of the order $2n$ in $d_\psi$,
 which is also of order one, (\ref{i6}).~\footnote{We now see that the cosmological `parameters' $\,\hep$ and $\het$ are of the second order in $d_\psi$ while the fundamental first-order parameter of our perturbation theory must naturally be $\chi_1 \equiv\,l^\pr$. } By substituting these operators into $\chi^2$ from (\ref{i5}) we find, after division by $\chi$, the \emph{exact relation}
 \be
  \chi = -l^\pr(\psi)\biggl(1-\frac{\chi^2}{6}\biggr)\biggl[1+\sum_2^\infty \frac{D_n \ast v(\psi)}{(-6)^{n-1}v\,l^\pr} \biggr]\,;\qquad \frac{D_2\ast v}{v\,l^\pr(\psi)} = \frac{\chi}{l^\pr}[\,l^{\pr\pr}+(l^\pr)^2] + (d_\psi\chi)\,,
 \label{i9}
 \ee
 where $D_n \ast v(\psi)$ denotes the action of the differential operator $D_n$ on $v(\psi)$, which is illustrated by the $D_2$ example (note that $\chi/l^\pr =-1 + O(\chi_1^2)$ and thus $\l^\pr$ in denominators is not dangerous). It is now easy to derive the second approximation $\chi_2$ by replacing in (\ref{i9}) $\chi$ by $\chi_1$ and neglecting the terms of the order $n\geq 3$:
 \be
  \chi_2 = -l^\pr(\psi)\biggl[1-\sixth (l^\pr)^2\biggr]\, \biggl\{1+\sixth[2 l^{\pr\pr}+(l^\pr)^2]\biggr\} = -l^\pr(\psi)\biggl[1+\third l^{\pr\pr} + O(\chi_1^4)\biggr]\,,
 \label{i10}
 \ee
 To calculate the third-order term we substitute $\chi_2$ for $\chi$ in (\ref{i9}) and take in the sum the terms up to $n=3$. This may be a simple exercise for a reader. \footnote{A simpler exercise is to show that the condition of linearity $\bv^{(2)}(\al)=0$,  with $\chi$ replaced by $\chi_1$, defines bilinear potentials $v(\psi)= v_0\,(\psi+\psi_0)^2$.}

 To illustrate perturbation theory we write the \emph{corrections to the inflationary parameters}, which follow from (\ref{i10}) for $\chi$ and $\bchi=1/\chi$. The corrected $\hep_1$ is $\hep_2 = \hep_1 [1+2\l^{\pr\pr}/3]$; for the simplest potentials $v=v_0\,\psi^{\,2N}$ we find in this approximation $\hep = \hep_1 [1-2\hep_1/3N+O(\hep_1^2)]$, where $\hep_1=\chi_1^2/2=2N^2/\psi^2$ and $|2\hep_1/3N|\,$ must be small. Not much more difficult is to find $\het_2$. From (\ref{i3}), (\ref{i6}-\ref{i7})  we see that $\het_1 = \l^{\pr\pr} +\,(l^\pr)^2$ and, using (\ref{i10}), we find
 \be
  \het_2 \,=\, \chi_2^2-d_\psi\,\chi_2 \,=\, \het_1 + (1/3)\,[\,2\,l^{(2)}\,\het_1 \,+\,l^{(1)} l^{(3)}-(l^{(2)})^2]\,.
 \label{i10a}
 \ee
 Taking for illustration the potential $v=v_0\,\psi^{\,2N}$ we easily find that
 \bdm
  \het_1 = 2N(2N-1)\,\psi^{-2}\,,\qquad \het_2 = \het_1\{\,1-\het_1\,[\,(4N-3)/\,3\,(2N-1)^2\,]\,\}\,,
 \edm
 which is $\het_2 = \het_1(\,1-\het_1/3)\,$ for $N=1$. With $\chi \cong \chi_2$, the number of $e$-folds is
 \be
  N_e = \int_{\psi_i}^{\psi_f} d\psi\,\bchi(\psi) =  \int_{\psi_f}^{\psi_i} \frac{d\psi}{l^\pr(\psi)}\biggl[1-\third l^{\pr\pr} + O(\chi_1^4)\biggr]= N_e^{(0)}(\psi) + \third \ln[\,l^\pr(\psi)]^f_i\,,
 \label{i10b}
 \ee
 where the first term is the standard result, and $N_e = [\,\psi^2/4N +\ln \psi/3\,]^i_f$ when $v=v_0\psi^{2N}$.

 Now, using Eq.(\ref{i9}), we can approximate equation (\ref{d10}) in the inflationary domain  $z^2\gg1$,
 \be
  dz/dx = (z^2 - 1) [z\,u(x) + 1] = (z^2 - 1)\,[\,u^\pr(x) +...] \,.,
 \label{i11}
 \ee
 see notation in (\ref{d10}). Then, recalling (\ref{d11})-(\ref{d12}) and taking $z(x) = -\coth(y)$, we find that
 \be
  dy/dx = 1 - u(x)\,\coth(y) = \,u^\pr(x) +...\,;\quad z\cong \coth(\chi_1/2) = -\coth u(x) \,,
 \label{i12}
 \ee
 where we have chosen an arbitrary integration constant zero (to have $z\rightarrow\infty$ when $u\rightarrow0$). Of course, this relation does not tell us anything new: returning to $\chi$ and $\psi$ we simply find our first-order relation $\chi\cong-\sqrt6\,\tanh (\,l^\pr/\sqrt6) \cong -l^\pr(\psi)$. However, this observation may signal that a perturbation series for the inflationary solutions can somehow be derived directly in the $\psi$-version, most probably, for special solutions like (\ref{d15}), (\ref{d15a}). It is instructive to look at a simplest \emph{asymptotic expansion} of $y(x)$ in powers of $1/x\sim u(x)\sim l^\pr$ for $v=v_0 x^{2N}$:
 \be
  y(x)=\sum_0^\infty y_{2n+1} \,x^{-(2n+1)}\,;\,\,\, \quad y_1 = N\,, \quad y_3 = \third N^3 - N^2,...\,,
 \label{i13}
 \ee
 which is obviously similar to inflationary perturbation expansions. This approach can probably be generalized to potentials for which $u(x)$ is expandable in powers of $1/x$. Similar asymptotic expansions can be used as approximations  of $y(x)$ near the finite-range singularities like $u(x)\sim1/(x-x_i)$ produced by the poles of the Higgs-type potentials $v(x)$. For inflationary potentials depending on powers of $e^{-gx}$, for which $\,u(x) = \sum_1^\infty u_n \,e^{-ngx}\,$, we can analogously exploit the expansions $\,y = \sum_1^\infty y_n \,e^{-ngx}\,$. 

 Note that the first-order terms of \emph{our approximations to the exact inflationary $\al$-solutions} reproduces the essentially approximate formulae used in the \emph{inflationary models}, while our higher-order terms give significant corrections to them. We also stress that all three inflationary parameters can be expressed in terms of one fundamental function $\chi^2=2\hep$ which describes all properties of the model. The second parameter, $\het$, can be explicitly expressed in terms of $\chi^2$ by Eq.~(\ref{i3}), which shows that $\het$  must be small as far as $\chi$ and $d_\psi \chi\,$ are small. The parameter $N_e$ can be  expressed as an integral of $\bchi \equiv \chi^{-1}$ over an inflationary $\psi$-interval and also is produced by $\chi$. Of course, in the present-day observations one measures not the $\chi$-functions  but some averages of $\,\hep,\,\het\,$, which nevertheless provide us with an important information on physics of inflationary models encoded in $\chi$ function.

 Note finally that the $\al$-version of inflation gives a direct and smooth transition from the `quantum' domain $\al<0$ to the classical one, where $\al \gg 1$. This picture looks surprisingly simple and coherent although further work is needed to clarify its relation to the standard cosmology. Even more fascinating might be attempts to leak out of the classical to quantum domain using ideas as simple as the classical dynamical cosmology.

 \section{A short summary}
 Here we briefly summarize the main results and problems that should be clarified in future. \textbf{1}.~The \emph{cosmological dynamical equations} are formulated \emph{in different gauges and versions}. We illustrate relations between them  on simple solutions and by integrable models.
 \\
 \textbf{2}.~The \emph{general properties of gauge independent $\chi$-equations} (\ref{d5})-(\ref{d9}), describing the main $(\al,\psi)$ portraits of isotropic cosmologies, are established in $\al$ and $\psi$ versions.
 \\
 \textbf{3}.~Equations (\ref{d4a})-(\ref{d4b}) allow us to derive the complete solution if $\chi(\al)$ or $\bchi(\psi)$ are known. Taking into account equations (\ref{d4ab}) we can in addition derive $\bv(\al)$ or $v(\psi)$.
 \\
 \textbf{4}.~We discussed \emph{different ways to determine cosmologies not using potentials}. A most natural one seems to first derive  $\chi^2(\al)$ using (\ref{d4c}), with the Hubble function $\xi^2(\al)$  as an input.
 \\
 \textbf{5}.~Although the $\chi$-equations depend only on $v^\pr(\psi)/v(\psi)$ and are thus insensitive to the sign of $v(\psi)\equiv\bv(\al)$, \emph{this sign is critically important for global properties of the solutions}. From (\ref{d5b}), (\ref{d11}),  (\ref{i2}) it follows that the solutions in the intervals with $v(\psi)>0$ are isolated from those in the intervals with $v(\psi)<0$ and must be studied separately.
 \\
 \textbf{6}.~We mostly considered potentials not changing  the sign and studied in detail models with \emph{positive potentials for which inflationary scenarios are natural}. We also can use and actually used our solutions and their expansions near the points $\psi_0$ where $v(\psi_0)=0$ and thus $v^\pr(\psi)/v(\psi)$ behaves as $(\psi-\psi_0)^{-1}\rightarrow \pm \infty$. This is a problem in the $\psi$-version because $(\chi^2-6)\,$ may change the sign with the potential, as follows from (\ref{d11}), (\ref{i2}). But in the $\al$-version it is no problem at all,  as can be seen from from expression (\ref{d6}) for $\chi^2(\al)$.
 \\
 \textbf{7}.~Probably, the \emph{most important results} are presented in Section~4, where we have found the \emph{exact solution of all equations for arbitrary $\bv(\al)$ and $k$}. The necessary condition for inflation is $\chi^2(\al)<6$ ($\,6\bchi^2(\psi>1$). To derive from $\chi(\al)$ standard inflationary scenarios we first suppose that the spatial curvature vanishes, $k=0$. Then, by fixing the  arbitrary integration constant, $C_0 =0$, we preserve the $v$-scale invariance of inflationary solution $\chi$ and derive its expansion from Eq.(\ref{i9}) as a sum, the $n$-th term of which for  $n\geq1$ has the form:
 \bdm
 -l^\pr(\psi) \sum_k c_n(\,k_1,...,k_{2n})\prod_{i=1}^{2n} [\,l^{(i)}(\psi)\,]^{\,k_i}\,, \quad \textrm{where}\,\,\sum_i{i k_i} = 2n\,, \,\,\,k_i \geq 0\,.
 \edm
 This \emph{inflationary perturbation expansion} can be obtained by the well-defined recursive algebraic iterations and gives higher-order corrections to the inflationary parameters $\,\hep,\,\het\,, N_e$~.
 \\
 \textbf{8}.~When $v(\psi)<0$ and thus $\chi^2(\al)>6\,$, $\,6\bchi^2(\psi)<1$, it is also convenient to use expansions of $\bchi(\psi)$ when it is small or close to $1/6$. In the last case we have derived \emph{asymptotic approximation} (\ref{d14d})  for $\psi\rightarrow \infty$ valid \emph{for a broad class of potentials} $v(\psi)$. We should mention an interesting one-parameter class of `bouncing' solutions (see (\ref{d16}) and (\ref{d17})), which exist when $v^\pr(\psi)/v(\psi) \sim 1/\psi$, and a special solution (\ref{d15a}) that probably is a separatrix. The \emph{global picture of solutions} with such properties are of great interest for ekpyrotic-bouncing scenarios and \emph{must be studied in future}.

 I believe that \emph{visualization of these structures}, drawing the $(\al, \psi)$ portraits, and using perturbative expansions for concrete inflationary, ekpyrotic, bouncing and other, more strange isotropic cosmologies may stimulate their better theoretical understanding.

 \bigskip
 \bigskip 
 {\bf Acknowledgment}\\The author is grateful to J.~Halpern for unceasing support. \\
 Useful discussions with A.~Starobinsky are kindly acknowledged.

 \section{Appendices}
 \subsection{On general isotropic cosmologies}
 A fairly general dimensional reduction of the Einstein gravity coupled to a scalar field $\psi$, which gives all possible spherically symmetric  cosmologies is described in \cite{ATF3}, \cite{ATF}. Following this procedure we derive the effective  two-dimensional Lagrangian describing spherical static states,  cosmologies, and waves. The starting point is the two-dimensional metric of the spherically symmetric space-time (we usually denote $e^{2\beta} \equiv \f$ and call it the \emph{dilaton} field),
 \be
  ds_4^2 = e^{2\alpha} dr^2 + e^{2\beta} d\Omega^2 (\theta , \phi) -
 e^{2\gamma} dt^2 + 2e^{2\delta} dr dt \, ,
 \label{eq1}
 \ee
 where  $\alpha, \beta, \gamma, \delta$ depend on $t$, $r$ and $d\Omega^2 (\theta , \phi)$ is the metric on the 2-dimensional sphere $S^{(2)}$. Then  the two-dimensional reduction of the four-dimensional Einstein gravity coupled to a scalar field $\psi$ is well known (here the prime denotes differentiations in $r$ and the dot - in  $t$):
 \be
  \cL^{(2)} =
  e^{\alpha + 2\beta - \gamma} (\dpsi^2 - 2\dbe^2 - 4\dbe \dal) -
  e^{-\alpha + 2\beta +\gamma} (\psi'^2 - 2\beta'^2 - 4\beta'\gamma')- e^{\alpha + 2\beta + \gamma} V(\psi) +  2\bk \,e^{\alpha + \gamma}\,,
 \label{eq2}
 \ee
 where $\bk =0, \pm 1$ is the standard curvature parameter, which vanishes for spatially flat cosmologies. This Lagrangian is obtained from the complete four-dimensional one, (\ref{L1}), by omitting the last term in metric (\ref{eq1}) and extracting the total derivative terms in (\ref{eq2}):
 \bdm
  \Delta \cL^{(2)} =
  -2 \,[(e^{\ga})^{\prime} \,e^{2\beta - \al}  \,+\, (\,e^{2\beta})^{\prime} \,e^{\ga - \al}\,]^{\prime}
 \edm
 (the omitted time derivative part is derived by replacement $\dif_r  \leftrightarrow i\dif_t$ and $\al \leftrightarrow \ga$).

 Let us also recall that (\ref{eq2}) is a gauge theory with two constraints: the total energy and momentum vanish according to the equations of motion derived by variations in all the variables. The origin of the constraints can be related to independence of the Lagrangian of the derivatives $\dal$ and $\prga$. It follows that $e^{\al(t)}$ and $e^{\ga(r)}$ become \emph{Lagrangian multipliers} in the one-dimensional static and cosmological reductions, respectively. A more rigorous treatment requires applying the ADM Hamiltonian formulation, \cite{ADM}.

 Variations of this Lagrangian give all the equations of motion\footnote{The equations of motion are the standard Einstein equation in the spherical coordinates. We need not write any Lagrangians to apply to them further separations of variables. However, introducing effective Lagrangians and Hamiltonians is extremely convenient, even in the classical environment, and will become indispensable if we turn to quantizing them (e.g.,\cite{CAF}).} except one constraint,
 \be
  -{\dot{\beta}}^{\prime} - \dot{\beta} {\beta}^{\prime} +
  \dot{\alpha} {\beta}^{\prime} + \dot{\beta} {\gamma}^{\prime} \,\,
  = \,\, \half \,\dot{\psi} {\psi}^{\prime} ,
 \label{eq4}
 \ee
 which should be derived before we omit the $\delta$-term in the metric (taking the limit $\delta \ra -\infty$). All other equations of motion can be obtained from the effective Lagrangian (\ref{eq2}).

 Now, the distinction between \emph{static} and \emph{cosmological} solutions is in the dependence of their `matter' field $\psi$ on the space-time coordinates. We call \emph{static} the solution for which $\psi=\psi(r)$. If $\psi=\psi(t)$ we call the solution \emph{cosmological}. There also exist the \emph{wave-like} solutions for which $\al, \beta, \ga$ and $\psi$ may depend on linear combinations of $t$ and $r$ but we here do not discuss this possibility. For both static and cosmological solutions the gravitational  variables in general depend on $t$ and $r$. This is important for the embedding the solution into higher dimensional theory but here we may forget about the dependence of cosmological solutions on the space coordinate.

 To obtain the one-dimensional equations we make further reductions by separating $t$ and $r$. It is clear that to separate the variables $r$ and $t$ in the metric we should require that
 \be
  \alpha = \alpha_0(t) + \alpha_1(r) , \quad \beta = \beta_0(t) +
  \beta_1(r) , \quad \gamma = \gamma_0(t) + \gamma_1(r) , \,
 \label{eq5}
 \ee
 Inserting this into the equations of motion one can find the restrictions on the gravitational (and, possibly on the matter) variables that must be fulfilled. The details can be found in \cite{ATF3}, where one can find the complete list of the static and cosmological spherically symmetric solutions.  The naive cosmological reduction (that supposes all the fields to be independent of $r$) does not give the standard FLRW cosmology with one scalar.  As was shown in \cite{ATF3} (see also the earlier paper \cite{ATF1}), one of the possible systems of conditions for separating the variables in the Einstein equations or in the Lagrangian is the following
 \be
 \label{eq6}
  \dot{\alpha} = \dot{\beta} \,, \quad \gamma' = 0 \,, \quad
  {\beta_1}'' + \bk \,e^{-2 \beta_1} = 0 \,, \quad
  2{\beta_1}'' +\, 3 {\beta_1'}^2 - \bk \,e^{-2 \beta_1} \,=\, 3k \,,
 \ee
 where the first two follow from Eq.(\ref{eq5}). The constant $k$ in the equations for $\beta_1$  is proportional to the 3-curvature of the space-time, and the third equation in (\ref{eq6}) is the isotropy condition. Any homogeneous isotropic cosmology must satisfy all four conditions.\footnote{Note that we here neglect inessential constant factors and have chosen $\alpha_1 = \gamma_1 = 0$.} Both the homogeneity and isotropy conditions follow from one equation
 \be
 \label{eq6a}(94)
  {\beta_1'}^2 - \bk \,e^{-2 \beta_1} \,=\, k \,,
 \ee
 which can easily be solved for all values of the parameters. Using Eqs.(\ref{eq6}), (\ref{eq6a}) we get the standard effective Lagrangian (\ref{L2}).  We see that for naive reductions the isotropy conditions in (\ref{eq6}) can be satisfied only if $\bk = 0$ and that the first condition is not dictated by (\ref{eq4}). Therefore, naive reductions give, in general, homogeneous non-isotropic cosmologies.

 In case of $\bk =k =0$ we can show that in the arbitrary naive cosmology with $\bk=0$ there exists a class of `isotropic' solutions  satisfying the condition $\dsig(t) \equiv \dal(t)  -\dbe(t)  = \textrm{const}$. It is easiest to demonstrate this in the Hamiltonian gauge $\ga = \al+3\beta$ in which:
 \be
  \ddot{\sigma} \equiv \ddal - \ddbe = k\,e^{2(\al+\beta)}\,, \qquad  2\ddal = e^{2\al+4\beta}\,v(\psi)\,\qquad 2\ddpsi = -e^{2\al+4\beta}\,v^\pr(\psi)\,.
 \label{eq6b}
 \ee
 When $k=0$, there exists the integral of motion $\dot{\sigma} = C_0$ and thus $\beta = \al - C_0 t$. The solutions belonging to the class with $C_0 =0$ are isotropic. Moreover, if the potential is exponential, i.e. $v^\pr(\psi) = gv(\psi)$ there appears one more integral, $\dpsi + g\dal = C_1$, and the model becomes integrable. Then the equation for $\al(t)$ can be easily reduced to the Liouville equation and explicitly solved. The result can be presented in a gauge independent form $\psi = f(\al)$.\footnote{Using the two integrals we find the equation for $\al$ looking like $\dda = \exp[a(t) + bt]$. Denoting $a+bt \equiv x(t)$, we find the Liouville equation $\ddx=\exp x$. This allows to find $t(\al)$ and exclude $t$ from the second integral.}

 This model of a relation between isotropic and anisotropic cosmologies is, of course, unrealistic. Whether some anisotropic cosmologies may have physically interesting isotropic limits is an interesting problem which is not discussed in this paper. With an additional scalar field, an evolution to isotropy looks possible, but then one should consider the complete system of three equations discussed in \cite{ATF}. This problem requires a separate investigation.

 \subsection{Integrable example of equation for $\bchi(\psi)$}
 Here we find the potential for which the $k=0$ reduction of Eq.(\ref{d8})  can be exactly solved and derive the correspondent solution, the general structure of which turn out similar to that of the general $\al$-solution. We first find a special solution $z_a(x)$ of (\ref{d10}) by supposing that $u(x)\,z_a(x) = a-1$ where $a$ will be determined later and $z_a$ satisfy $z_a^\pr =a(z_a^2 -1)$. Obviously, $z_a = - \coth^{\ve} (ax)$ where $\ve = \pm1$ and thus $u(x) = (1-a)\tanh^{\ve} a \bx$, where $\bx\equiv x-x_0$. Putting from now on $x_0 =0$ we find the corresponding potential $\tv(x)$ (recall (\ref{d10})),
 \be
  \tv(x) = v_0 (\,|\,e^{ax}\,+\,\ve e^{-ax}\,|/2\,)^{\,2(1-a)/a}\,, \quad \ve = \pm1 \,.
 \label{b1}
 \ee
 By the way, $u(x) \equiv v^\pr(x)/2\,\tv(x)$ satisfies the simple equation
 \be
  u^\pr (x) = a(a-1)^{-1} [\,u^2 - (a-1)^2\,] \,.
 \label{b2}
 \ee
 Now, let us try to find the general solution by the substitution
 \be
  z =\,y +\,z_a =\,y + (a-1)/u \,=\,y - \coth^{\ve} (ax) \,.
 \label{b3}
 \ee
 It follows that $y(x)$ satisfy the Abel equation,
 \be
  y^\pr = u\,y^3 + (\,a + 2\,u\,z_a)\,y^2 + (2\,a\,z_a + u\,z_a^2 - u)\,y \,,
 \label{b4}
 \ee
 which can be explicitly integrated if $\,a + 2\,u\,z_a =0$; this is possible if $a=2/3$. Then,
 \be
  y^\pr = u\,y^3 - (u + 1/3u)\,y \,, \qquad  1/3u = \coth^{\ve}(2x/3)\,
 \label{b5}
 \ee
 and $w \equiv y^{-2}\,$ satisfies the linear equation:
 \be
  w^\pr(x) = 2(u + 1/3u)\,w -2u \,.
 \label{b6}
 \ee
 We write the general solution for $\ve =-1$ which is regular at $x=0$ ($a=2/3$),
 \be
  w(x) = \,\cosh^2 (ax)\,[\,\cosh(2ax) + C_0\, \sinh(2ax)\,]\,, \qquad z(x) = (w(x))^{-\half} \,+ \,\tanh (ax)\,.
 \label{b7}
 \ee
 The solution for $\ve=+1$ will be regular if we choose the minus sign of the root in (\ref{b7}) to cancel the singularity of $z_a = 1/3u(x)$, see (\ref{b1}). The function $w(x)$ is positive for all $x$ if $C_0^2 < 1$ and is equal to $\cosh^2 (2x/3)\exp(\pm 4x/3)$ for $C_0 =\pm1$. Similarly to the general solutions in the $\al$-picture, we have here the positivity problem and the second sheet also emerges if $C_0 <-1$. In any case, by integrating $z(x)$ we can derive the explicit expression for $\al(\psi)$ that, in principle can be inverted to obtain $\psi(\al)$. Thus $\psi(\al)$ and $\bv(\al)\equiv v[\psi(\al)]$ will implicitly depend on two arbitrary parameters -- one is $C_0$ and the second is the integration constant $\psi_0$ emerging in deriving $\psi(\al)$.

 \subsection{On $v(\psi)$ versus $\bv(\al)$ in simple solutions}
 Here we consider important examples of deriving the potential $v$ both in $\al$ and $\psi$ pictures. We discuss in some detail the problem of finding potentials for which there exist some simple solutions of equations (\ref{H1a})-(\ref{Ha}), where we omit bars and denote $\tau$-derivatives by dots:
 \be
  2\,\dxi + \eta^2 \,+\, 2k e^{-2\al} = 0\,, \quad 2\,\deta\,+\,6\,\eta\,\xi \,+\,\prv(\psi)\,=\, 0\,,
 \label{H1b}
 \ee
 \be
  v(\psi) = 6\,\xi^2 - \eta^2 -6\,k\,e^{-2\al}\,.
 \label{Hb}
 \ee
 The simplest and physically interesting solutions constructed on one of two ad hoc guesses: 1.~$\dxi = C_0$ or 2.~$\deta=C_0$. The key idea is to solve the potential-independent equation, to derive $\chi(\al)$ or $\bchi(\psi)$ giving $\al(\psi$), and then to use constraint (\ref{Hb}) for finding $v(\psi)$. The aim is twofold: first to learn something on a rather nontrivial relation between the standard and $\al$ `versions' and, second, to find simplest `dual' potentials $v(\psi)$ and $\bv(\al)$.  The result of these simplest, almost trivial considerations looks somewhat unexpected, especially, in case of non-vanishing curvature term $ke^{-2\al}$.

 We consider in some detail only the first, simpler solutions. When $C_0=0$ we easily find
 \be
  \dal(\tau) \equiv \xi = \xi_0\,, \quad \al(\tau) = \xi_0\,(\tau-\tau_0)\,,  \quad \eta=k_0\, e^{-\al(\tau)}\,,   \quad (k_0^2\equiv -2k)\,.
 \label{c1}
 \ee
 The partial map $\chi(\al)$ and the corresponding $\psi(\al)$ are very simple in this case,\footnote{In fact, we find only the part of the complete portrait that corresponds to the chosen solution. As is demonstrated in the main text, we can reconstruct the complete portrait for any potential $\bv(\al)$ in the $\al$ picture but even in this simplest example we do not know the portrait for the corresponding potential $v(\psi)$.}
 \be
  \chi(\al) = \frac{d\psi}{d\al} = \frac{\eta}{\xi} = \frac{k_0}{\xi_0}\,e^{-\al}\,; \quad \tpsi \equiv (\psi - \psi_0) = \int\chi(\al) = -\frac{k_0}{\xi_0}\, e^{-\al}\,; \quad \bchi(\psi) = -\frac{1}{\tpsi} \,.
 \label{c2}
 \ee
 It follows that the potential in $\al$ and in $\psi$ versions is very simple,
 \be
  \bv(\al) \,=\, 6\,\xi_0^2 \,+\, 2\,k_0^2\,e^{-2\al}\,=\, \bv\,[\al(\psi)] \,=\,6\,\xi_0^2\,+\, 2\,\xi_0^2\,(\psi - \psi_0)^2 = v(\psi)\,.
 \label{c3}
 \ee
 It is not difficult to check that all dynamical equation are satisfied in all versions. Note that the arbitrary parameters have different meaning -- $\xi_0$ is the Hubble constant while integration constant $\psi_0$ is unimportant as $\psi$ is in fact defined up a shift.

 The case $\dxi = C_0 \equiv -\eta_0^2/2$ is solved quite similarly and we skip details. We have:
 \be
  \eta^2 = \eta_0^2 + k_0^2\, e^{-2\al}\,, \quad \xi = C_0 (\tau - \tau_0)\,, \quad  \al = C_0\,(\tau - \tau_0)^2 /2 + \al_0 \,, \quad \xi = -\eta_0\,\sqrt{\al_0 - \al} \,\,.
 \label{c5}
 \ee
 Now it is easy to derive the portrait and the simple expression for its limit for $k_0 \ra 0$,
 \be
  \tpsi\,= \int \chi(\al)\, = \int d\al\,(1 + k_0^2 \,\eta_0^{-2}\,e^{-2\al})^{1/2} \,(\al_0 - \al)^{-1/2} \,\ra 2\sqrt{\al_0 - \al}\,,
 \label{c6}
 \ee
  and easily find $\bv(\al)$ and $v(\psi)$; for the last we only write $k_0 \ra 0$ limit:
 \be
  \bv(\al) = 6\,\eta_0^2\,(\al_0 - \al) - \eta_0^2 + 2k_0^2\,e^{-2\al} = \bv[\al(\psi)] \ra (3/2)\,\eta_0^2\,(\psi - \psi_0)^2  - \eta_0^2 \,.
 \label{c7}
 \ee
 This limit gives the same  potential $\sim\psi^2$ obtained for the $\dxi=0$ example and this also is true for the $\deta = 0$ case. The condition $k=0$ significantly simplifies solutions, especially, in the $\al$-version (see Section~4). Equations (\ref{c6})-(\ref{c7}) with $k_0 \neq 0$ and similar ones for the case $\deta = C_0 \neq 0$ give more complex potentials. Though the problems are of physics interest we cannot them in the present paper.

 It would be of interest to find and study potentials giving physically interesting solutions \emph{and at the same time} analytically accessible in the $\al$-picture. Such an approach looks viable and deserving careful elaboration but it will require much deeper understanding of the analytical structure and the precise meaning of $\psi(\al)$ as well as of the $\al$-version as a whole. For these reasons, we are compelled to leave this task to future investigations. At the moment this problem is solved in the frame of inflationary perturbation theory, Section~4. As a next step in this direction, one may analyze from this point of view the examples of this Appendix and explicitly integrable models briefly discussed in the main text of this paper.

  \newpage

 \end{document}